\def\bra#1{\mathopen{\langle#1\,|}}
\def\ket#1{\mathclose{|\,#1\rangle}}
\def\braket#1#2{\mathopen{\langle#1\,|}\mathclose{\,#2\rangle}}
\def\sumint{{\textstyle \sum_{ \xi}}\hspace{-0.52cm} 
         {\int_{\cal C}} \hspace{0.14cm}}

\documentstyle[prd,aps,eqsecnum]{revtex}
\title{\bf The Quantum Aspects of Relativistic Fermion Systems\\
           with Particle Condensation}
\author{S. Ying}
\address{Physics Department, Fudan University\\
Shanghai 200433, People's Republic of China}

\draft

\begin{document}
\maketitle
\begin{abstract}
  A consistent local approach to the study of interacting relativistic
fermion systems with a condensation of bare particles in its ground or
vacuum state, which may has a finite matter density, is developed. The
attention is payed to some of the not so well explored quantum aspects
that survive the thermodynamic limit. A 4-vector local field, called
the primary statistical gauge field, and a statistical blocking
parameter are introduced for a consistent treatment of the
problem. The effects of random fluctuations of the fields on local
observables are discussed. It is found that quasiparticle
contributions are not sufficient to saturate local observables.  The
property of the primary statistical gauge field are discussed in some
detail. Two models for the strong interaction are then introduced and
studied using the general framework developed. Four possible phases
for these models are found. The possibility of spontaneous CP
violation and local fermion creation in two of the four phases is
revealed. The implications of the finding on our understanding of some
of the strong interaction processes are discussed.
\end{abstract}

\tableofcontents

\section{Introduction}
\label{sec:Intro}

   Fermions are the fundamental building blocks of the observable
universe, which, at certain level, is supposed to be understoodable in
terms of quantum field theories (QFTs) like QCD, standard model of
electroweak interacting, etc.. They are,
however, less familiar to us theoretically compared to bosons partly
due to their lack of classical correspondences. An understanding of
Nature therefore requires a more direct understanding of the behavior
of fermions.

Represented by anticommuting Grassmann numbers in the path integration
formulation of a QFT, fermions are harder to handle in numerical
simulations (like the lattice ones) than bosons. It is therefore
desirable to integrate the fermionic degrees of freedom out (in the
path integration sense) analytically. This turns out to be an easy task,
at least formally, in most of the cases which deals with a Lagrangian
density that is only quadratic in the fermion fields. The result is
usually a much more complicated effective action for the bosonic
fields of the system to be functionally integrated by various means
like a numerical calculation or simulation, an approximated analytic
computation, modeling and possibly the mixture of all of them. The
fermion loop effects in numerical simulations are in some sense less
well understood than their bosonic counterparts in a theory at
present. The problems become more sever in the presence of a finite
chemical potential in lattice simulations \cite{LQCDmu}. This
situation calls for more analytic efforts to understand the effects of
fermionic quantum fluctuations in an interacting system since it
indicates that our understanding of the problem is still insufficient.

The traditional treatment of the finite density problems (at finite
temperature) in statistical mechanics is based upon the grand
canonic ensemble in which the partition function is $Z= Tr
e^{-\beta(\widehat H - \mu_{ch}\widehat N)}$ with $\beta$ the inversed
temperature, $\widehat H$ the Hamiltonian and $\widehat N$ the
particle number of the system. A global chemical potential $\mu_{ch}$
is introduced to select, among all possible particle numbers, the
corresponding set of particle numbers that are different from each
other by a finite quantity in the thermodynamic limit. Since only
those extensive quantities that are proportional to the (infinite)
volume of the system are relevant, the above-mentioned differences are
irrelevant to bulk thermodynamic quantities. This makes the grand
canonic ensemble equivalent to the canonic ensemble \cite{Huang}
in which the number of particles are kept fixed. The usually
calculated quantities, called the {\em apparent particle number}
here, are expressed as $\overline N_{app} =\beta^{-1}\partial \ln Z /
\partial \mu_{ch}$ is formally identical to what is called the {\em
absolute particle number} $\overline N_{abs} = \int d^3 x Tr \widehat
\rho(\mbox{\boldmath{$x$}},t=0) e^{-\beta(\widehat H-\mu_{ch}\widehat
N)}/Z$. It can be realized that the identification between $\overline
N_{app}$ and $\overline N_{abs}$ is not mathematically warranted in
the thermodynamic limit since the particle number $\widehat N$ is a
macroscopic operator with eigenvalues proportional to the volume of
the system. we thus expect that the formal equivalence between
$\overline N_{app}$ and $\overline N_{abs}$ and many other physical
observables so computed to be broken down under certain conditions
especially when the quantum fluctuation effects are taken into
account. In order to characterize the deviation of the quantity
$\overline N_{app}$ from $\overline N_{abs}$, a {\em dark component}
for each physical local observables in a relativistic QFT is
introduced.  For example, the dark component of the fermion number
density operator is defined as $\overline {\Delta \rho} = (\overline
N_{abs} - \overline N_{app})/\Omega$ with $\Omega$ the volume of the
system. The questions to be assessed are whether or not $\overline
{\Delta \rho} = 0$ and what is the origin of the dark component when
$\overline {\Delta \rho}\ne 0$.

  Such a possibility is important to study because some of the
questions posted for the vacuum state of a relativistic system
governed by certain QFT are different from the ones
asked for the condensed matter system in which the quantities under
study like the average particle number density is finite with its
{\em absolute} values playing no direct physical role in physical processes
and in which the spacetime resolution (energy) of the observation is
usually low.  The vacuum state of a relativistic system, especially
the trivial one, is characterized by its nothingness nature, namely,
all physical observables in the trivial vacuum state are by definition
zero. In the non-trivial vacua of the system, certain physical
quantities develop finite values which need to be evaluated
correctly. Those quantities like the fermion (baryon) number density
and associated energy density should in principle manifest themselves
in gravitational process at the macroscopic level.  In addition,
global quantities have no direct physical meaning in a classical
relativistic system according to principle of relativity.  It is
expected that the apparent quantities like $\widehat N_{app}$ are not
sufficient ones in the study of the vacuum state of a relativistic
system.  Rather, one should go back to the absolute quantities like
$N_{abs}$ defined above. Therefore, for a better marriage between
relativity and quantum mechanics, instead of a global chemical
potential $\mu_{ch}$, a localized quantity called the {\em primary
statistical gauge field} $\mu^\alpha(x)$ seems to be necessary, which
leads to a local theory for the finite density problem. The functional
derivative of the logarithm of the new partition functional with
respect to $\mu^\alpha(x)$ at certain spacetime location does give the
absolute particle number density since it is a finite number.

The principle of {\em locality} has far reaching consequences in the
development of modern physics. Implied in Maxwell's equations for
electrodynamics, it motivated the birth of relativity in which it is
raised to a principle that governs all physical laws in classical
physics.  Localities in quantum field theories are implemented in most
of the theories regarded as fundamental like the quantum
electrodynamics and quantized non-abelian Yang-Mills gauge theories of
the standard model.  Locality in the quantum field theories, including
the fundamental ones, for non-vanishing matter and energy density is
not, as a matter of fact, fully implemented.  Such theories contain
inconsistencies, at least at the conceptual level, that have to be
removed. 

Locality of a symmetry generates a gauge one. For the $U(1)$ invarince
corresponding to the conservations of fermion number, a new local
symmetry called the {\em statistical gauge invariance} with
the gauge field $\mu^\alpha(x)$ is induced after its localization.

The introduction of a primary statistical gauge field is expected to
produce a series of new problems and opportunities.  One of the goals
of this paper is to solve these problems and to explore such
opportunities so that to develop a consistent framework beyond the
quasiparticle picture using which the problems related to the vacuum
state of a relativistic fermion system can be systematically tackled.

   One of the most important problems in understanding a system
governed by a QFT that contains interaction is to determine the phase
structure of its vacuum. The vacuum is a state of the system that has
the lowest energy that can be different from the trivial one for
interacting systems.  The non-trivial vacuum of an interacting system
covered by this study are the ones that contain macroscopic
condensation of particles in various form.  It has zero overlap with
the corresponding trivial one when the thermodynamic limit is taken. Such
a state can not be reached by a perturbative computation, which is
local in nature and can only change finite number of particles. A
large set of non-trivial vacua of interest are expected to be
describable in terms of a set of parameters that characterize the
macroscopic condensation of bare particles. Some of such parameters
are called the order parameters since they are indicators of a
spontaneous breaking down of certain global symmetries of the
system. They are accompanied by massless Goldstone bosons that
generate long range orders which stabilizes the symmetry breaking
states. Others, together with the order parameters, specify the
macroscopic condensation of bare particles in the non-trivial vacuum
in a more detailed way. For example, the vacuum of the light quark
system (up and down quarks) is known to be condensed with
quark--antiquark pairs. This phenomenon, which induces the spontaneous
breaking down of a chiral symmetry, is shown to happen in the lattice
QCD simulation \cite{LatticQCD} and is supported by experimental facts
due to the success of the partial conservation of axial vector current
(PCAC) relationship and resulting current algebra \cite{PCACsuc}.

  For a system with a large mass gap (compared to the typical
excitation mass scale of the system), the condensation of bare
particles is unlikely to occur in its vacuum state. So, for the
purpose of this paper, I consider light (compared to the typical
excitation mass scale of the system) fermion systems. They can be
approximately represented by a massless fermion system.

 An interaction amongst massless fermions and antifermions of the
right sign and magnitude is expected to generate a non-vanishing
number of fermion--antifermion pairs from the bare vacuum. Under
certain conditions, the number of such pairs can become macroscopic
(or proportional to the volume of the system) in the thermodynamic
limit. In such a case, it is expected to has a phase transition. Such
a phase transition was show to happen to the Nambu Jona--Lasinio (NJL)
model in the quasiparticle approximation (the meaning of which is
going to be specified in the following sections).  It is also shown in
Ref. \cite{Ying1,Ying11} that fermion pairs (and antifermion pairs)
can condense to lead to a different phase ($\beta$ phase) that belongs
to the same chiral symmetry breaking chain, namely $SU(2)_L\times
SU(2)_R\to SU(2)_V$.

 In general, the condensation of fermion pairs in the vacuum of an
interacting fermionic system belongs to one of three categories: 1)
condensation of correlated {\em fermion and antifermion} pairs 2)
condensation of correlated {\em fermion pairs} (and possibly some
correlated {\em antifermion pairs} 3) condensation of correlated {\em
antifermion pairs} (and possibly some correlated {\em fermion pairs}).
The spin, flavor and color combination of the condensed pairs
determines the nature of the symmetry breaking channel of the
non-trivial vacuum on a finer basis. Since a macroscopic number of
fermions and antifermions are pumped out of the bare vacuum in the
non-trivial vacuum, it is natural to ask what are the effects on the
its energy density by considering not only the contributions from the
long distance or/and time interval correlated quasiparticle
excitations, but also, at least partially, the contributions from some
of the transient and short distance quantum fluctuations within the
system. To investigate the later effects within a consistent QFT, 
certain new theoretical concepts and tools in describing
and interpreting the related physics, to which this work put its
emphasis on, is proven necessary.

  The paper is divided into three major parts. The first part consists
of sections \ref{sec:Intro} and \ref{sec:Summ}, which gives an
introduction and a summary. In the second part, which consists of
sections \ref{sec:General}, \ref{sec:Instab} and \ref{sec:FDth}, a
consistent general approach to the relativistic fermionic system at
both zero and finite density is developed.  The third part includes
sections \ref{sec:Models} and \ref{sec:Vacua}, in which two 4--fermion
interaction models for the strong interaction are introduced and
studied using the method developed in part two; some novel properties
of these models are revealed using the new tools.

The more detailed arrangement of the paper is given in the following.
In section \ref{sec:General}, the general framework used to handle the
fermionic system is discussed. An 8--component ``real'' representation
for the fermion fields is adopted. The distinction between the
Minkowski and Euclidean spacetime formulation of the problem is
emphasized.  It is pointed out that in the Euclidean spacetime
formulation of the problem, additional contributions due to certain
quantum penetration of field configurations to classically forbidden
region, to which quasiparticle approximation in Minkowski spacetime
can not access, can be included in the effective action. I also
motivate the need for a distinction between local and global
observables in a relativistic QFT. The existence of the dark component
is demonstrated.  Section \ref{sec:Instab} is devoted a tentative
local approach to the finite density problem for a Lagrangian density
that conserves the fermion number. Such a formalism is used to study
the question of whether or not a state with non-vanishing fermion
number density can has a lower energy (density) when the vacuum of the
system is in a phase different from the trivial one. Spontaneous CP
violating stationary points in the Euclidean spacetime is shown to
exist in a phase, called the $\alpha$ phase of a massless fermion
system, with non-vanishing $\overline\psi\psi$ vacuum expectation
value.  It is argued that such a result is at least physically not
acceptable. Based on these findings, a consistent theory is developed
in section \ref{sec:FDth} for a relativistic fermion system in which a
phase transition with particle condensation has occurred.  A
statistical blocking parameter $\epsilon$ is introduced. The question
of the statistical gauge invariance of the system due to the original
global $U(1)$ symmetry related to fermion number conservation is
addressed.  To demonstrate the relevance of the theory, two models for
strong interaction are introduced in section
\ref{sec:Models}. Their phase structures are then studied. The vacuum
of both of them has three different phases. One is the trivial (bare)
vacuum, which is called the $O$ phase; the second is the above
mentioned $\alpha$ phase with fermion and anti-fermion pairs
condensation; the third kind of phases, called the $\beta$ phase and
$\omega$ phase respectively, are phases that spontaneously break the
global $U(1)$ symmetry related to the fermion number because of a
condensation of fermion pairs and antifermion pairs.  These phases are
further analyzed in section \ref{sec:Vacua} by using the formalism
developed in section \ref{sec:FDth}. A spontaneous CP violation is
found to be present in the $\beta$ and $\omega$ phases. Also, the
spontaneous creation of matter, which is relevant to Cosmology, is
shown to appear naturally in the $\beta$ and $\omega$ phases.
Finally, the main results are summarized and discussed in section
\ref{sec:Summ} which also contains an outlook.

\section{General Discussions}
\label{sec:General}

\subsection{Fermion representation}

    The fermion field is represented by an 8--component ``real''
spinor $\Psi$. It is essential for a consistent formulation of a
relativistic finite density field theory in a functional (or path
integration) approach developed in this work \cite{TFTQFT} (see also
Appendix \ref{app:FF}) and for other aspects of the QFTs
\cite{Larsen}. In this representation, $\Psi$ satisfies the {\em
reality} condition
\begin{eqnarray}
   \overline\Psi(x) &=&\Psi^T(-x) \Omega_0\label{Psi}, \label{PsiBar}
\end{eqnarray}
where the $\Omega_0$ matrix is
\begin{eqnarray}
\Omega_0&=& O_1 C = \left (\begin{array}{cc} 0 & - C^{-1} \\ C &
                  0\end{array}\right )
\label{G}
\end{eqnarray}
with $C$ the charge conjugation operator. The matrices $O_1$ is one of
the three Pauli matrices $O_{1,2,3}$ acting on the upper and lower 4
components of $\Psi$.

For the case of the number of flavors $n_f$ for the fermions is less
than three, the symmetry transformation of the 8 component $\Psi$
can be made to be the same as the 4 component representation $\psi$ by
imposing a slightly different ``reality' condition on $\Psi$ 
given by Eq. \ref{PsiBar} with $\Omega_0$ replaced by
\begin{eqnarray}
\Omega&=&  \left (\begin{array}{cc} 0 & - C^{-
1} \rho^{-1} \\ C \rho &
                  0\end{array}\right ).
\label{G1}
\end{eqnarray}
Here $\rho=1$ if $n_f=1$ and $\rho= i\tau_2$ if $n_f=2$. 
More detailed properties of this representation for the case of $n_f=2$ 
are discussed in Refs. \cite{Ying1,Ying11,TFTQFT,SSSconf}. 
They will not be repeated here.
 
If $n_f\ge 3$, the reality condition Eq. \ref{G1} is no longer 
valid. In such a case, the general ``reality'' condition is given by the 
original Eq. \ref{PsiBar} and the representation of $\Psi$ under flavor 
$SU(n_f)$ transformation generated by
\begin{eqnarray}
     T_a &=& \left \{ \begin{array}{cc} 
                         t_a O_3 & \hspace{1.5cm}\mbox{If $t_a$ is
                           symmetric}\\
                         t_a     & \hspace{1.8cm}\mbox{If $t_a$ is
                           antisymmetric}                          
                      \end{array} \right .
\end{eqnarray}
have to be adopted. Here $a=1,2,\ldots,n_f^2-1$ 
and $t_a$ is the generator for the symmetry
transformation in the 4 component representation of $\psi$. The set of
matrices $T_a$ and
$t_a$ ($a=1,2,\ldots,n_f$) belong to equivalent adjoint representation
of flavor $SU(n_f)$ transformation. 

  It should be emphasized that the intention of introducing an
8--component representation for the fermions is different from
the one related to the doubling of degrees of freedom in the
closed time path approach \cite{CTPT,CTPT1} to
non-equilibrium (also equilibrium)
problems or thermal field dynamical \cite{TFD} approach to equilibrium
problems in two aspects: 1) this work is devoted to the study of zero
temperature physics where, in the language of the closed time path approach,
all the dynamical fields considered in this work lies on a single time
axis running from negative infinity to positive infinity rather than
on two time axis that form a closed loop
2) there is no doubling of degrees of freedom for
the fermions here since the constraint Eq. \ref{PsiBar} is
systematically implemented in the formalism. In case the present
formalism is to be extended to finite temperature \cite{TFTQFT} 
or to the
non-equilibrium situations, a doubling of the components of the
8--component ``real'' representation has to be made in addition.

\subsection{Minkowski spacetime formulation}

The generating functional can be written as
\begin{eqnarray}
      e^{W[J,\overline\eta,\eta]} &=& \int \prod_i D[f_i] D[\Psi]
                   e^{i\int d^4x \left ({1\over 2} \overline\Psi i
                   S_F^{-1}[f]\Psi + {\cal L}_B[f] + \overline\Psi\eta
                   + \overline\eta\Psi + \sum_k J_k f_k\right)},
\label{General_W}
\end{eqnarray}
where $J=\{J_1,J_2,\ldots,J_n\}$ are a collection of external fields
coupled to the corresponding boson fields $f=\{f_1,f_2,\ldots,f_n\}$,
${\cal L}_B[f]$ is the Lagrangian density for the boson fields $f$,
and $\eta$, $\overline\eta$ are external Grassmann fields coupled to
the fermion fields $\overline\Psi$, $\Psi$.  Here $f_i$ can be either
real or complex.  $W[J,\overline\eta,\eta]$ generates the Green
functions of the fermion or the boson fields. The possible gauge
fixing conditions, which can be implemented by multiplying $\prod_i
D[f_i]$ a set of $\delta$ functions or by introducing ghost fields
\cite{GaugeTheory}, are unimportant to this work and are suppressed in
the sequel.

The bosonic part of the Lagrangian density will not be
specified in this investigation. This allows the results of this work
to be useful in a wide class of problems.
For example, in case of QCD, the boson fields are
8 gluon fields ${\cal B}_a^\mu(x)$ $(a=1,2,\ldots,8)$ with the full Lagrangian
density in
the 8 component representation for quarks provided in
Appendix \ref{app:QCD}.

In the study of the vacuum properties, fermion degrees of
freedom can be eliminated first
\begin{eqnarray}
      e^{W[J,\overline\eta,\eta]} &=&  \int \prod_i D[f_i]
                   e^{{1\over 2}LnDet \gamma_0 i S_F^{-1}[f] +
                    {1\over 2} \overline\eta S_F[f] \eta
                     + i\int d^4x\left ({\cal L}_B[f] +
                     \sum_k J_k f_k\right )}.
\label{Eff_W1}
\end{eqnarray}
Here ``$Det$'' denotes functional determinant.
The proper vertex generating functional for the boson fields is then
\begin{eqnarray}
     \Gamma[f] = W[J,0,0]-i\sum_k J_k f_k.
\label{Vertex}
\end{eqnarray}
The stationary configuration $f$ determined by the equation
\begin{eqnarray}
     {\delta \Gamma[f]\over \delta f_i} &=& 0\hspace{0.7cm}(i=1,2,\ldots,n)
\label{EqofMotion}
\end{eqnarray}
with $f_i$ ($i=1,2,\ldots,n$) spacetime independent
determines the phase structure of the vacuum. In general,
$\Gamma[f]$ is difficult to compute directly.
It is useful to define an effective action $S_{eff}$ for the boson
fields as
\begin{eqnarray}
   S_{eff}[f] &=& -i{1\over 2}LnDet \gamma_0 iS_F^{-1}[f] + i{1\over 2}LnDet
   \gamma_0 iS_F^{-1}[0] + \int d^4 x {\cal L}_B[f] \nonumber\\
      &=& - i{1\over 2} Sp Ln S_F^{-1}[f] +
            i{1\over 2} Sp Ln S_F^{-1}[0] + \int d^4 x {\cal L}_B[f],
\label{Seff}
\end{eqnarray}
where $Sp$ denotes the functional trace and
a constant (infinite) term independent of the boson fields
are subtracted. Eq. \ref{Eff_W1} becomes
\begin{eqnarray}
      e^{W[J,\overline\eta,\eta]} &=&  \int \prod_i D[f_i]
                   e^{i S_{eff}[f] +
                    {1\over 2} \overline\eta S_F[f] \eta
                     + i\int d^4x \sum_k J_k f_k}.
\label{Eff_W2}
\end{eqnarray}

Starting from $S_{eff}[f]$, either systematic improvement beyond the
Gaussian approximation or numerical simulations like lattice
computation can be made. A more detailed discussion of the local
quantum fluctuations around the mean field is given in Appendix
\ref{app:thedark}.  A formal relation between $\Gamma[f]$ and
$S_{eff}[f]$ is developed in Ref. \cite{Jackiw} and discussed in
Appendix \ref{app:CJT}.  When the fluctuation in $f$ is only treated
at an one loop level, the vertex functional $\Gamma[f]$ is (see
Appendix \ref{app:CJT})
\begin{eqnarray}
 \Gamma[f] &=& iS_{eff}[f] - {1\over 2} Sp Ln D G^{-1}[f]
,\label{1-loop-gamma} 
\end{eqnarray}
where $D$ is the bare propagator for $f$ and $\delta^2 S_{eff}/\delta
f\delta f$ is symbolically denoted as $G^{-1}[f]$ .
Under such an
approximation, the solution to the equation
\begin{eqnarray}
      {\delta S_{eff}[f]\over \delta f_j} + {i\over 2} Sp G[f]
        {\delta G^{-1}[f] \over \delta f_j} &=& 0
      \hspace{0.7cm}(j=1,2,\ldots,n)
\label{Seff-stable}
\end{eqnarray}
determines the vacuum phase structure of the system. It will not be
further discussed in this paper since the results of this paper depend
only on some of the bosonic fields $\{ f_i \}$ as a solution to
Eq. \ref{EqofMotion} being different from zero.

The effective action $S_{eff}[f]$ can be expressed in terms of the
spectra of the operator $\gamma_0 iS^{-1}_F[f]$, which is Hermitian. The
eigenequation of interest is
\begin{eqnarray}
       \gamma^0 iS^{-1}_F[f] \Psi_\lambda = \lambda[f] \Psi_\lambda
\label{EigenEq}
\end{eqnarray}
with $\Psi_\lambda$ the eigenvector. In the time independent case,
$S_{eff}[f]$ in terms of $\lambda$ is 
\begin{eqnarray}
    S_{eff}[f] &=& -i{T\over 2} \sumint {dp^0\over 2\pi}
    \ln{\lambda_{p^0,\xi}[f]\over \lambda_{p^0,\xi}[0]} +
    \int d^4 x {\cal L}_B[f]
,\label{Seff2}
\end{eqnarray}
where $T$ is the temporal dimension of the system ($T\to\infty$),
$p^0$ represents the energy of the eigenvector $\Psi_\lambda$ and
$\xi$ is a collection of other quantum numbers that completely
determines a single eigenvector $\Psi_\lambda$. The order in which
$p^0$ integration and $\xi$ summation is carried out is important in
general since they both are divergent before the subtraction. The
symbol $$\displaystyle \sumint {dp^0\over 2\pi}(\ldots)$$ is
understood as that neither the sum over $\xi$ nor the $p^0$
integration is not done first but they are done in a covariant way.
The order between them depends on situations which are discussed in
the following and in Appendix \ref{app:FF}.

Due to the logarithmic function, the integrand in the complex $p^0$
plane is multivalued, the physical contour ${\cal C}$ with
Feynman--Mathews--Salam causal structure \cite{TFTQFT} for the $p^0$
integration is shown in Fig. \ref{Fig:ConCon1}. The conventional
computation of the energy density of the vacuum with a non-covariant
cutoff, which is called the {\em quasiparticle approximation}, can be
represented by a distortion of contour to curve I of Fig.
\ref{Fig:ConCl2}. It corresponds to a summation of the energies of
individual stationary quasiparticle orbits in the negative energy
Dirac sea with a fixed (time-independent) background configuration for
$f$.

\subsection{Euclidean spacetime formulation}

The path integration representation of the generating functional
$W[J,\overline\eta,\eta]$ on the right hand side of Eq. \ref{General_W}
contains ambiguities associated with the non-specification of the
initial and final field configurations. For the transition amplitude
between a given pair of initial and
finial field configurations, the contributing intermediate states can
in principle be different from the vacuum state interested here. The lowest
energy configuration corresponding to the vacuum is automatically
projected out in an Euclidean spacetime computation with a sufficiently
large Euclidean time for a large set of
proper initial and final field configurations.

In addition, the Minkowski effective action Eq. \ref{Seff} has extrema
(or is stable) only at configurations of the boson fields that are of
classical nature. An important class of configurations corresponding
to pure quantum mechanical effects, namely the tunneling effects
through potential barrier or penetration into the classically
forbidden field configurations are expected to be missing in a
steepest--descent or Gaussian approximation. The Euclidean action
obtained by replacing time variable $t$ by $-it$ (with $i^2=-1$) can
be stable at either the time independent configurations, which are the
same ones as in the Minkowski approach, or the configurations that
correspond to the tunneling or penetration that are quantum mechanical
in origin. It is important to realize that the later stable
configurations are always absent in the set of stable configurations
of the Minkowski effective action despite they constitute an important
contribution to the energy density of the system. The Euclidean
spacetime formulation has been used to find important stable field
configurations called instantons in non-Abelian gauge theories.  They
correspond to stable finite action field configurations due to the
tunneling between gauge field configurations of different winding
numbers. Other applications of finding the tunneling effects by using
the Euclidean spacetime formulation and their connection to the WKB
method in quantum mechanics can be found in e.g. Ref. \cite{Eucl-App}.
It will not be elaborated here. The effects that are interested in
this study are the ones that survive the thermodynamic limit and are
thus non-finite action effects despite they can be decomposed into a
collection of random finite action ones (see Appendix
\ref{app:thedark}).

The generating functional for the Green functions in the Euclidean
spacetime formulation can be expressed as
\begin{eqnarray}
      e^{W[J,\overline\eta,\eta]} &=&  \int \prod_i D[f_i]
                   e^{ S_{eff}^E[f] +
                    {1\over 2} \overline\eta S^E_F[f] \eta
                     + \int d^4x \sum_k J_k f_k}
\label{Eff_W3}
\end{eqnarray}
which also serves as a definition of the Euclidean effective action 
$S_{eff}^E[f]$. The Euclidean propagator $S^E_F[f]$ is found by using
the rules discussed in the following.

For a fermion system, the simple replacement $t\to -it$ is not
sufficient to obtain a consistent Euclidean spacetime formalism from
the corresponding Minkowski one due to the fact that a fermion is not
a scalar particle. More sophisticated set of transformations are
needed.  Formally, an Euclidean effective action for fermions can be
obtained by making a continuous change of the metric
\cite{Eucl-Ferm}. The result of change for a Dirac particle can be
summarized by
\begin{eqnarray}
   g^E_{\mu\nu} = \{-,-,-,-\},\hspace{0.3cm}
   \gamma_0^E = i\gamma^5,&\hspace{0.2cm} &
   \gamma_5^E = -i\gamma_0,\hspace{0.3cm}\gamma^E_i = \gamma_i
\label{Eucl-Rules}
\end{eqnarray}
and the effective action Eq. \ref{Seff} in the Euclidean spacetime 
formulation becomes
\begin{eqnarray}
    S^E_{eff}[f]  &=& {1\over 2} \left
               \{SpLn \left [i\gamma_5 (S^E_F[f])^{-1}\right]
                 -SpLn\left[ i\gamma_5 (S^E_F[0])^{-1}\right]
               \right \} + \int d^4x_E {\cal L}_B[f^E],\label{E-Seff}
\end{eqnarray}
where the Euclidean propagator $S^E_F[f]$ is also obtained by using
substitution rules listed in Eqs. \ref{Eucl-Rules}.  In terms of the
eigenvalue of the hermitian operator $i\gamma_5 (S^E_F[f])^{-1}$ that
satisfies
\begin{eqnarray}
   i\gamma_5(S^E_F[f])^{-1} \Psi_{\lambda} &=& \lambda[f] \Psi_{\lambda}
.\label{Eucl-EigEq}
\end{eqnarray}
The Euclidean correspondence of Eq. \ref{Seff2} is then
\begin{eqnarray}
    S_{eff}[f] = -i{T\over 2} \sumint {dp^0\over 2\pi}
    \ln{\lambda_{p^0,\xi}[f]\over \lambda_{p^0,\xi}[0]} +
    \int d^4 x {\cal L}_B[f]
,\label{E-Seff2}
\end{eqnarray}
where the superscript ``$E$'' on top of a quantity in Euclidean
spacetime is suppressed in there and in the following whenever no
confusion is thought to occur.  It can be demonstrated that
Eq. \ref{E-Seff2} can be obtained from Eq.  \ref{Seff2} by a
distortion of the $p^0$ integration contour to curve II shown in
Fig. \ref{Fig:ConCl2}.

\subsection{Local and global observables} 

 The physical observables in a classical theory consistent with
relativity are local ones like the charge density, energy density
etc.. Unlike in the non-relativistic world, the global observables
like the total charge and the energy of the system are not directly
measurable. In a given frame, however, the global observables can be
defined as a spatial integration of the local observables on an equal
time hypersurface in the Minkowski spacetime. The meaning of the
global observables can only be defined operationally. The particular
value of a global observable has to be determined by the following
procedure.  First a synchronization of the clocks of a group of
observers on each spacetime point should be carried out, then let each
observer measure the corresponding density at his/her location in
spacetime at the same time and finally sum (integrate) each observers
finding to obtain the value of the global observable.

  The global observables in quantum theories, including the
relativistic QFTs, are regarded as physical
observables.  Let us consider certain density observable $\widehat
\rho(x)$ in a QFT.  The corresponding global
observable $\widehat Q$ is defined simply as
\begin{eqnarray}
   \widehat Q &=& \int_{\Sigma} d^3x \widehat \rho(x)
\end{eqnarray}
with $\Sigma$ the equal-time hypersurface in certain reference
frame. Then the matrix elements of $\widehat\rho(x)$ and $\widehat Q$,
namely, $\bra{f}\widehat\rho(x)\ket{i}$ and $\bra{f}\widehat Q\ket{i}$
define the corresponding observables.

  Since both the local observable $\widehat\rho(x)$ and the
classically not defined global one $\widehat Q$ are defined in a
relativistic QFT, one can naturally ask the following
question, namely, is there any difference between
$O=\bra{\Omega}\widehat Q\ket{\Omega}$ and $O'= \int d^3 x
\bra{\Omega}\widehat\rho(x)\ket{\Omega}$ measured in a state $\Omega$?
It can be shown that this is a relevant question for a relativistic QFT.

A measurement of the total charge or charge density of a system
normally consists of supplying an external global field or local field
coupled to the corresponding observables in a known way
and then measure the {\em response of the system} like the force that
the external field exerts on the system or the amount of increase of
some proper defined ``potential'' of the system, which allows the
observer to deduce the total charge or charge density measured.  Let
us consider the measurements, in the above sense, of the global and
the local observables in the vacuum state of the system expressed
symbolically as
\begin{eqnarray}
       O_0 &=& \lim_{J\to 0} 
           \bra{0_J}\widehat Q\ket{0_J}= \lim_{J\to 0}
           \bra{0_J}\int_{\Sigma} d^3x 
               \widehat \rho(x) \ket{0_J},\label{O0}\\
      O'_0 &=& \int_{\Sigma} d^3x \lim_{\delta j(x)\to 0} 
                     \bra{0_{\delta j(x)}} 
                            \widehat \rho(x) \ket{0_{\delta j(x)}}  
               ,\label{Op0}
\end{eqnarray}
where $J$ is a global external field coupled to $\widehat Q$ and
$\delta j(x)$ is a local external field taking non-vanishing value
only at $x$ that couples to $\widehat \rho(x)$ and $\ket{0_J}$ and
$\ket{0_{\delta j(x)}}$ are the corresponding vacuum states.  The
first one $O_0$ corresponds to a measurement of $\widehat Q$ directly;
the second one corresponds to the integration of a (infinite) set of
measurements of $\rho(x)$ on a space-like hypersurface.  These two
measurements can in principle be different in a system governed by a
QFT because of the random local quantum fluctuations
of the fields that are not suppressed in the thermodynamic limit.

Such a potential difference guarantees the possible existence of the
dark component in a QFT.  The fact that localized
random quantum fluctuations are not suppressed in an interacting
QFT, especially in some of the non-trivial phases of
the system, are discussed in more details in Appendix
\ref{app:thedark} where it is also shown that the conventionally used
quasiparticle picture in many-body theory and QFT is not sufficient to
saturate the local observables due to the existence of the dark
component. Such a deviation from the conventional physical picture
based upon quasiparticles gets less and less significant as the
resolution of our observation respect to spacetime gets lower and
lower compared to the typical size of the localized random
fluctuations $f_a$ of the system. In such a case these fluctuations
are more and more suppressed and the contribution of the dark
component becomes smaller and smaller resulting in 1) an emergency of
a quasiparticle dominated picture for the system and 2) the validity of
the results obtained based upon a global chemical potential $\mu_{ch}$
in the grand canonic ensemble in low resolution (energy) observations.

\section{A Naive Local Quantum Finite Density Field Theory and Its Problems}
\label{sec:Instab}

\subsection{A tentative formalism for the local finite density
  fermionic field theory}

     In order to develop a theoretical framework consistent with the
requirement of locality, a new quantity called primary statistical
gauge field $\mu^\alpha(x)$ is introduced in the following. The
motivation for its introduction is discussed in the introduction and
in the above sections.  The other reason that it is treated as a local
variable is due to the fact that total fermion number has limited
meaning in a relativistic system that are generated in the past not
infinite long ago. The total number of fermions in such a system is
not an observable since there are regions outside the horizon that are
classically non-detectable even in principle due to the constraint of
causality.

\subsubsection{The asymptotic grand canonic ensemble}

   The generating functional corresponding to Eq. \ref{General_W}
for a fermionic system with finite density can be formally written as
\begin{eqnarray}
     e^{W[J,\overline\eta,\eta,\mu]} &=& \int \prod_i D[f_i] D[\Psi]
                   e^{i\int d^4x \left ({1\over 2} \overline\Psi i S_F^{-1}[f]\Psi
                      + \mu_\alpha j^\alpha +
                   {\cal L}_B[f] + \ldots \right )}
,\label{General_W_FD}
\end{eqnarray}
where ``$\ldots$'' denotes the source terms and the fermion number
current $j^\mu(x)$ is
\begin{eqnarray}
     j^\mu(x) &=& {1\over 2}\overline\Psi(x) \gamma^\mu O_3 \Psi(x).
\label{fermion-curr}
\end{eqnarray}
In the Minkowski spacetime, quantity on the left hand side of the
above equation is the transition amplitude of the system from properly
weighted initial
field configurations $\phi_i$ at $t=-\infty$ to final field
configurations $\phi_f$ at $t=+\infty$, namely,
\begin{eqnarray}
     e^{W[J,\overline\eta,\eta,\mu]}
                  &=&\sum_{\{\phi_i,\phi_f\}} {\cal W}[\phi_f,\phi_i] 
                              \braket{\phi_f,t=+\infty}{\phi_i,t=-\infty},
\end{eqnarray}
where ${\cal W}[\phi_f,\phi_i]$ is the weight functional 
discussed in the following.

 These initial and final fields are
considered free fields. The interaction terms are adiabatically switched on
at certain large negative time $-T$ and switched off at
certain large positive time $T$. Such a technical manipulation does
not affect the actual local physical observables like the energy
density, fermion number density, etc. at time $t$ that is far
from both $-T$ and $T$ due to locality.

  Since the particle content for free fields at $t=\pm\infty$ is
clear, which allows a straight forward statistical interpretation in
terms of number of particles in each single particle state of the
system. It is natural to assume that the initial (final) state are in
the grand canonic ensemble with the weight ${\cal W}[\phi_f,\phi_i]$
for the summation determined by the factor $\lim_{\beta\to\infty}
\exp[-\beta(E_0-\mu N)]$\footnote{We are interested in the zero
temperature case here. The form of the weight functional for free
fields at finite temperature can be found using the method given in,
e.g., Refs.  \cite{CTPT,CTPT1}, which involves two time axis: one runs
from $-\infty$ to $+\infty$ on real $t$ axis; the other lies below it
on the complex $t$ plan. In the zero temperature limit, only the
contributions from the real $t$ axis is nonzero, which leads to
Eq. \ref{General_W_FD}. }, where $E_0$ is the total energy and $N$ is
the total fermion number of the (free) system. $\mu$ agrees with the
time component of the spacetime independent part of the primary
statistical gauge field $\mu^\alpha(x)$. Such an ensemble is called
the {\em asymptotic grand canonic ensemble} here.

The asymptotic 
grand canonic ensemble differs from the grand canonic ensemble
for it allows for a local approach to the relativistic 
finite density problem and 
the existence of the dark component discussed in the
above sections and in Appendix \ref{app:thedark}. Its predictions, however, approaches that of the
grand canonic ensemble in the non-relativistic situations in which
the spacetime resolution or energy in a
measurement is low. This point is demonstrated in Appendix
\ref{app:thedark} at the mean field level.

Two points must be pointed out. The first one is that although the
time interval $2T$ between the (adiabatic) switching on and off of the
interaction terms are let to go to infinity in the final result, the 
thermodynamic limit, in which the spacetime box that contains the
system approaches infinity, is taken first. Therefore $T$ is still
infinitesimal compared to the temporal size of the system. The second
one is that, as a result, for the conserved quantities, 
like the total fermion number, the extreme value of it 
picked out \cite{Huang} in the asymptotic grand canonic ensemble with a fixed
value of $\mu$ is unmodified. The later point is elaborated in the following.

\subsubsection{The fermion number density in the asymptotic grand
               canonic ensemble}

Before continuing the development of the formalism, it is important to
reveal a property of the asymptotic grand canonic ensemble related to
the $U(1)$ symmetry corresponding to the fermion number conservation.
The conservation of $j^\alpha(x)$ due to the $U(1)$ symmetry can be
derived from the Noether's theorem at the classical level, namely,
\begin{eqnarray}
   \partial_\mu j^\mu(x) &=& 0.
\label{jB-cons}
\end{eqnarray}
Those interaction Lagrangian densities for which Eq. \ref{jB-cons}
remains true at the quantum level are considered despite the fact that
this symmetry is not explicitly related to a gauge symmetry for which a
superselection sector in the Hilbert space of the system exists. The
reason to consider such a class of models is because for a quark
system, the fermion number is identical to the baryon number, which is
conserved to a high precision due to the lack of any convincing
evidence of the proton decay in observation at the present.

For an uniform system, Eq. \ref{jB-cons} implies that
\begin{eqnarray}
      \left .
      {\partial \overline\rho\over \partial g_i}\right |_{\mu^\alpha}
                                   & = & 0 \hspace{1cm}(i=1,2,\ldots)
\label{p-rho-pgi}
\end{eqnarray}
in the asymptotic grand canonic ensemble with $\{g_i\}$ representing a
set of interaction coupling constants and the derivative taken by
keeping $\mu^\alpha$ unchanged. For an uniform system, the mean local
fermion density is a function both of the coupling constants $\{ g_i
\}$ and $\mu^\alpha$, which is now spacetime
independent. Eq. \ref{p-rho-pgi} implies that {\em the mean local fermion
density of the system $\overline\rho$ is only a function of
$\mu^\alpha$, it is independent of the interaction coupling constants
$\{g_i\}$.} This property allows us to find out the relationship
between $\overline\rho$ and $\mu^\alpha$ easily by considering a
non-interacting system. It is discussed in Appendix \ref{app:FF}. The
result for a massless fermion with $n_f$ flavors and $n_c$ colors is
\begin{eqnarray}
    \overline\rho &=& {n_f  n_c\over  3 \pi^2} \mu^3
,\label{bar-rho-mu}
\end{eqnarray}
where $\mu \equiv \sqrt{\mu^2}$. The simplicity of
Eq. \ref{bar-rho-mu} in the asymptotic grand canonic ensemble is due
to the fact that the interaction Lagrangian density conserves fermion
number. It implies that for an interacting massless fermion system,
local quantity $\overline\rho$ is non-zero as long as $\mu$ is
non-zero even when a phase transition in its vacuum state that
generates a finite gap for the lowest excitation of the system has
happened. Such a qualitative behavior is required following the
discussion given in Appendix \ref{app:thedark}.  Eq. \ref{bar-rho-mu}
is however an exact relation in the asymptotic grand canonic
ensemble.

   After an exploration of the implications of the $U(1)$ symmetry of
the class of models under consideration in the asymptotic grand
canonic ensemble, we are in a position to further
develop the formalism for the investigation of the density fluctuations of
the vacuum after a phase transition. Since Eq. \ref{General_W_FD} can be 
obtained from Eq. \ref{General_W} by the replacement
\begin{eqnarray}
i\rlap\slash\partial&\to&
i\rlap\slash\partial+\rlap\slash\mu O_3
\label{part-repl}
\end{eqnarray}
the effective action $S_{eff}$ of the system is changed to
\begin{eqnarray}
    S_{eff}[f,\mu] 
    &=& -i{T\over 2} \left (\sumint {dp^0\over 2\pi}
    \ln{\lambda_{p^0,\xi}[f;\mu]\over \lambda_{p^0,\xi}[0,\mu]}
    + \sum_{\xi} \int_{\cal C} {dp^0\over 2\pi}
       \ln{\lambda_{p^0,\xi}[0;\mu]\over \lambda_{p^0,\xi}[0,0]}
    \right )
     +\int d^4 x  {\cal L}_B[f]
\label{Seff2-FD}
\end{eqnarray}
with the eigenvalues $\lambda$ obtained from the
following equation
\begin{eqnarray}
       \gamma^0 iS^{-1}_F[f;\mu]
                   \Psi_\lambda = \lambda[f;\mu] \Psi_\lambda
,\label{EigenEq2}
\end{eqnarray}
where $S_F^{-1}[f;\mu]$ is derived from the corresponding one
in Eq. \ref{EigenEq} by making the substitution Eq. \ref{part-repl}.
Since the question interested in this study is related to a 
comparison of the energies of states with different fermion densities when the
interaction is present, I consider the following effective action
obtained from $S_{eff}$ by a Legendre transformation
\begin{eqnarray}
       \widetilde S_{eff} &=& S_{eff} - \int d^4 x \mu_\alpha \overline j^\alpha.
\label{tS_eff}
\end{eqnarray}
It is a canonic functional of $\overline j^\alpha$ with $\mu^\alpha$
implicitly depending on it.  For an uniform system like the vacuum,
the second logarithmic term in Eq.  \ref{Seff2-FD} is calculated in
such a way as not to over-count the already known (and included)
quantum fluctuations of the free fields. The result is given in
Appendix \ref{app:FF} (Eq. \ref{W00-3}), it is
\begin{eqnarray}
-i{T\over 2} \sum_{\xi} \int_{\cal C} {dp^0\over 2\pi}
\ln{\lambda_{p^0,\xi}[0;\mu]\over \lambda_{p^0,\xi}[0,0]} &=&
\int d^4 x \left (\mu \overline \rho - \overline \varepsilon\right )
\end{eqnarray}
in the rest frame of the density.
Together with Eqs. \ref{bar-rho-mu}, Eq. \ref{tS_eff} for an uniform
system takes the form
\begin{eqnarray}
   \widetilde S_{eff} &=&
-i{T\over 2} \sumint {dp^0\over 2\pi}
    \ln{\lambda_{p^0,\xi}[f;\mu]\over \lambda_{p^0,\xi}[0,\mu]}
     -{n_f  n_c\over 4\pi^2} \int d^4 x\mu^4
     +\int d^4 x {\cal L}_B[f].
\label{Seff2-FD2}
\end{eqnarray}
The effective potential to be minimized for an uniform system
is then defined by
\begin{eqnarray}
   V_{eff} &=& - \lim_{V_3T\to\infty} \widetilde S_{eff}/V_3T
     \nonumber \\ &=&
\lim_{V_3\to\infty}
i{1\over 2V_3} \sumint {dp^0\over 2\pi}
    \ln{\lambda_{p^0,\xi}[f;\mu]\over \lambda_{p^0,\xi}[0,\mu]}
     +{n_f  n_c\over 4\pi^2} \mu^4
     - {\cal L}_B[f]
,\label{Veff-FD}
\end{eqnarray}
where volume $V_3 = L^3$ with $L$ the spatial dimension of the system.

\subsection{Euclidean Instability of the $\alpha$ Phase and Its CP Problem}

The $\alpha$ phase for a massless fermion system can be shown to
realize by using the Nambu Jona--Lasinio (NJL) model \cite{NJL}. In
the 8--dimensional ``real'' representation for the fermion
spinor, the NJL
model with 2 flavors ($n_f=2$) and 3 colors ($n_c=3$) are given in 
appendix \ref{app:NJL}. The NJL model is studied extensively in
the literature by
assuming $\mu^\alpha=0$ in the vacuum so far. It's possible that such a
plausible assumption is in fact incorrect.

    With Eq. \ref{Veff-FD}, the question of whether or not the
$\alpha$ phase is stable against fluctuations in $\mu^\alpha$ 
can be studied. In the $\alpha$ phase, Eq. \ref{Veff-FD} takes the
following form
\begin{eqnarray}
    V_{eff} &=& 6i\int_{\cal C} {d^4p\over (2\pi)^4} \ln\left (
            1- {\sigma^2\over p_+^2} \right ) \left (
            1- {\sigma^2\over p_-^2} \right )
               + {1\over 4 G_0}\sigma^2 +  {3\over 2\pi^2} \mu^4
,\label{Veff-a-FD}
\end{eqnarray}
where $p_+^\mu = (p^0+\mu^0,\mbox{\boldmath $p$}+\mbox{\boldmath
  $\mu$})$ and
$p_-^\mu= (p^0-\mu^0,\mbox{\boldmath $p$}-\mbox{\boldmath $\mu$})$.

\subsubsection{Quasiparticle approximation and its problem}

  In the quasiparticle approximation, the $p^0$ integration contour is
the one shown in Fig. \ref{Fig:ConCl2}. The localized quantum
fluctuations of the order parameter $\sigma$ are not included
following such a contour.  With the help of Appendix \ref{app:FF}, Eq.
\ref{Veff-a-FD} can be shown to have the following form
\begin{eqnarray}
       V_{eff} = {3\over 2\pi^2} \mu^4 +  {1\over 4 G_0} \sigma^4
                 - {6\over \pi^2} \int_{k_F}^{{\Lambda_3}}
                 d|\mbox{\boldmath $p$}| |\mbox{\boldmath $p$}|^2
                 \left ( \sqrt{|\mbox{\boldmath $p$}|^2+\sigma^2} -
                 |\mbox{\boldmath $p$}| \right )
,\label{Veff-FD-2}
\end{eqnarray}
where $k_F = \theta(\mu-\sigma)\sqrt{\mu^2-\sigma^2}$ with $\theta(x)$
the step function and
${\Lambda_3}$ is the cutoff in the 3--momentum that defines the
model. It is a monotonic increasing function of $\mu$. So the
stationary point for $V_{eff}$ is at the position $\mu=0$, as
expected.

 There is however an inconsistency related to the fermion number in
the quasiparticle approximation to the $\alpha$ phase.  Eq.
\ref{bar-rho-mu} holds for any phase of a model Lagrangian density for
which the fermion number is conserved. On
the other hand, the average fermion number for an uniform system in
the quasiparticle approximation is
\begin{eqnarray}
     \overline n = {2\over \pi^2}\theta(\mu-\sigma) (\mu^2-\sigma^2)^{3/2}
\label{semi-c-n}
\end{eqnarray}
which differs from Eq. \ref{bar-rho-mu} and the results of Appendix
\ref{app:thedark} even qualitatively. The reason, as is mentioned
generally in section \ref{sec:General}, is because the quasiparticle
approximation is related to a Gaussian approximation to the generating
functional of the model in the Minkowski spacetime formulation. Such
an approximation can only take into account of the contributions from
propagating modes or quasiparticles. The quasiparticles are not
fundamental building blocks, namely, the bare particles, of the
system; they are composite objects representing coherent excitations
of infinite pairs of bare particles and anti-particles. In the phase
that quasiparticles propagate, the excitations correspond to the bare
particles of the system damp in spacetime when produced.  They are
not, however, absent in a time interval that is sufficiently
short. They form an important component, which is called the dark
component (see Appendix \ref{app:thedark}), that contributes to the
fermion number density and other relevant local physical
quantities. The contribution of such a dark component is expected to
be taken into account, at least partially, if we formulate our
semi-classical approximation in the Euclidean spacetime to sample the
important contributions of pure quantum mechanical configurations.

\subsubsection{Euclidean stationary points and the CP problem}

   In the Euclidean spacetime, Eq. \ref{Veff-a-FD} can be evaluated by using 
the $p^0$ integration contour shown in Fig. \ref{Fig:ConCl2} together
with a {\em covariant} cutoff. This
operation corresponds to the replacement rule given by Eq.
\ref{Eucl-Rules} plus $\mu^0\to -i\mu^0$, which is equivalent to the
change $p^0\to ip^0$ and $\mu^0 \to \mu^0$ in the expression for the
Minkowski 
effective action after the tracing over spin, isospin and color degrees
of freedom is taken. The time component of $\mu^\alpha$ is kept
unchanged since it is regarded as an external field and is related to the
fermion density given by Eq. \ref{bar-rho-mu}.
The result is
\begin{eqnarray}
    V_{eff}/\Lambda^4 = -{3\over \pi^3} \int^1_0 dy y^3 \int_0^1 dx
           \sqrt{1-x^2}
           \ln{(y^2-\widetilde\mu^2+\widetilde\sigma^2)^2 + 4\widetilde\mu^2
             y^2 x^2 \over (y^2-\widetilde \mu^2)^2 + 4\widetilde\mu^2 y^2
             x^2} + {1\over 16\pi \alpha_0}\widetilde\sigma^2
             + {3\over 2\pi^2} \widetilde \mu^4
\label{Eucl-Veff-FD}
\end{eqnarray}
$\widetilde
\sigma \equiv \sigma/\Lambda$, $\widetilde \mu \equiv \mu/\Lambda$,
$\alpha_0=G_0\Lambda^2/4\pi$ and with $\Lambda$ the covariant
cutoff in the Euclidean momentum space. 

In the $\alpha$ phase with non-zero $\sigma$, the dependence of
$V_{eff}/\Lambda^4$ on $\mu$ is plotted in Fig. \ref{Fig:a-phase-mu}.
The minima of $V_{eff}$ is not located at $\mu=0$ for any finite
$\sigma$ but some finite $0<|\mu_0|<\sigma$. This is physically not acceptable.

There are two problems related to such a vacuum with non-vanishing
fermion density if the dominate phase of the
physical strong interaction vacuum is considered to be in the $\alpha$ phase.

The first
one is related to the strong CP problem. Since in such a background field as
$\mu^\alpha$, the two CP conjugate eigenstates of a neutral particle has
different energies; it implies the occupation number for
one eigenstate can be larger than the other in physical processes at
sufficiently low energy. This in turn will cause much too large CP violation
phenomena not observed in nature. Let us consider the only
system, namely the $K^0/\overline K^0$ system, in which a CP violation was
observed. A mechanism for the
explanation of the CP violation in the $K^0/\overline K^0$ system was
proposed long ago \cite{TDLee}. Its basic idea is to assume a $\mu^\alpha$
like potential through out the space
that differentiates $K^0$ and $\overline K^0$. However, such a $\mu^0$ is
estimated to have a value of order $10^{-8}$ eV. It is much smaller than
the average value found here, which is of the order of $10^2$ MeV if
$\Lambda$ is taken to be $1$ GeV.

The second one is related to the dark matter problem. If there exists
an uniform $\mu^0\sim 10^2$ MeV field in the universe at the present,
the dark component of the baryon number density would be of order of the
nuclear matter density, which is much larger than what it is expected.
Had such a scenario been  correct, the universe would not has lived
until today.

Facing these two serious problems, a solution is needed to be looked for. There
are at least two alternatives.
The first one, which is likely to be correct, is that there are
something missing in the computation procedure used so far.
The second one is that our notion that the dominate
phase of the universe is in the $\alpha$ phase is wrong. The later alternative is
unlikely to be correct since there is a large body of empirical facts
that support such a notion.

With this consideration in mind, I turn next to an investigation of the
first alternative.

\subsection{A Fock space inspection
               of the vacuum structure of the $\alpha$ phase and the
               blocking effects}

 It is known that in the $\alpha$ phase of the NJL
model treated in the mean field approximation, the vacuum can be related to
the bare one by an unitary transformation before the thermodynamic
limit $L^3\to\infty$ is taken. It can be explicitly written as
\begin{eqnarray}
     \ket{\mbox{vac}} = \prod_{{\bf p},h} e^{i\widehat O({\bf p},h)} \ket{0}
\label{V-BV}
\end{eqnarray}
with ${\bf p}$ the three momentum,
$h$ the helicity label and operator $\widehat O$ expressed in term
of the creation operators ($a^\dagger_{{\bf p}h}$, $b^\dagger_{-{\bf
    p}h}$) and
annihilation operators ($a_{{\bf p}h}$, $b_{-{\bf
    p}\lambda}$) of the bare fermions as
\begin{eqnarray}
       \widehat O({\bf p},h) &=& {i\over 2}\theta_{\bf p}
       \left [ a^\dagger_{{\bf p}h}
       b^\dagger_{-{\bf p}h} -
       b_{-{\bf p}h} a_{{\bf p}h} \right ],
\label{O-oper}
\end{eqnarray}
where $\cos\theta_{\bf p}=|{\bf p}|/E_{\bf p}$ and $E_{\bf p} =
\sqrt{|{\bf p}|^2 + \sigma^2}$. The annihilation operators
for the quasiparticles $\alpha_{{\bf p}h}$ and the antiquasiparticle 
$\beta_{{\bf p}h}$
of the system are related to the bare ones
through a Bogoliubov transformation
\begin{eqnarray}
    \alpha_{{\bf p}h} &=& \cos({1\over 2}\theta_{\bf p})
                    a_{{\bf p}h} - \sin({1\over 2}\theta_{\bf
                      p}) b^{\dagger}_{-{\bf p}h},
                    \label{alpha_oper}\\
    \beta_{{\bf p}h} &=& \sin({1\over 2}\theta_{\bf p})
        a^\dagger_{-{\bf p}h} + \cos({1\over 2}\theta_{\bf p})
        b_{{\bf p}h}.\label{beta_oper}
\end{eqnarray}

{}From Eqs. \ref{V-BV}, \ref{O-oper} two properties can be seen: 1)
the $\alpha$ phase vacuum is a superposition of states in the Fock
space of bare fermions with increasing number of fermion and
anti-fermion pairs and 2) after the thermodynamic limit
$L^3\to\infty$ is taken the overlap between the bare vacuum $\ket{0}$
and the $\alpha$ phase vacuum $\ket{\mbox{vac}}$ tends to zero. This
implies 1) the occupation of fermions and anti-fermions in the
$\alpha$ phase vacuum can have blocking effects on the creation
operation of bare particle states and 2) the true $\alpha$ phase
vacuum can not be reached by perturbative iterations starting from
some preassumed state of the system without introducing some kind of
macroscopic variables into the path integration formalism.

\section{A Consistent Local Quantum Finite Density Field Theory}
\label{sec:FDth}

\subsection{The statistical blocking parameter}

\subsubsection{The motivating Fock space study}

    I develop a theory that takes into account both the
fermion--antifermion pair condensation and the resulting blocking
effects in this section. As mentioned in the previous section, the
bare vacuum with zero pairs of fermion and antifermions is not a
suitable initial state that starts the path integration computation
since it has zero overlap with the true vacuum state of the system
after the phase transition.  Instead, the starting state is of the
following kind
\begin{eqnarray}
    \ket{\phi_0} &=& \sum_{n,\xi} C^n_\xi \ket{ (f\overline f)_\xi^n}
                  = \sum_{n,\xi}C^n_\xi \ket{n\xi}
\label{init-state}
\end{eqnarray}
with $n$ the number of the fermion--antifermion pairs and $\xi$
other quantum numbers that completely specify the state.
The diagonal matrix element of the
evolution operator of the system can be expressed in terms of path
integration over the dynamical fields of the system
\begin{eqnarray}
        \braket{\phi_0,t=+\infty}{\phi_0,t=-\infty} &=&
        \lim_{\begin{array}{c} t_f\to\infty\\t_i\to-\infty \end{array}}
         \bra{\phi_0} e^{-i\widehat H (t_f-t_i)} \ket{\phi_0}\nonumber \\
         &=& \sum_n \sum_{\xi\xi'} C^n_\xi
         C^{n *}_{\xi'} \braket{n\xi',t=+\infty}{n\xi,t=-\infty} + \ldots
,\label{evolution1}
\end{eqnarray}
where $\widehat H$ is the total Hamiltonian of the system and $\ldots$
represents the off diagonal contributions to the transition
amplitude  between states
with different pairs of fermion and antifermion. For a discussion of
the eigenstate of the total Hamiltonian like the vacuum state, the
off diagonal contributions with macroscopically different initial and
final fermion--antifermion pairs are expected to be 
effectively absent in the final result after  the thermodynamic
limit.

The transition amplitude $\braket{n\xi',t_f}{n\xi,t_i}$ with the external
fields present is then written in
terms of path integration, namely,
\begin{eqnarray}
      \braket{n\xi',t=+\infty}{n\xi,t=-\infty} &=&
            N \int D[\Psi]\prod_i D[f_i]
            e^{i\int d^4x \left ({1\over 2} \overline\Psi i S_F^{-1}[f]\Psi
                     + {\cal L}_B[f] + \overline\Psi\eta + \overline\eta\Psi + 
                     \sum_k J_k f_k\right)},
\label{General_W_FDB}
\end{eqnarray}
where $N$ is a constant. The formal manipulations, which express the
above functional integration over fermion degrees of freedom by an
effective action $S_{eff}$, remain mostly unchanged, except $S_{eff}$
depends now on a new statistical parameter $\epsilon$
\begin{eqnarray}
      \braket{n\xi',t=+\infty}{n\xi,t=-\infty} &=&
            N' \int \prod_i D[f_i]
                   e^{iS_{eff}[f,\mu,\epsilon] +
                    {1\over 2} \overline\eta S_F[f] \eta
                     + i\int d^4x \sum_k J_k f_k}.
\label{Eff_W_FDB}
\end{eqnarray}
For a stationary situation, the effective action $S_{eff}[f,\mu,\epsilon]$ is
given
by Eq. \ref{Seff2}. The constraint that both of the initial and
the final states considered are configurations with both of the bare fermion
and antifermion states below energy $\epsilon$ filled is
implemented by a distortion of the contour for $p^0$ integration from the
one in Fig. \ref{Fig:ConCon1} to that of Fig. \ref{Fig:ConFDB0}.

In the thermodynamic limit of $L^3\to \infty$, the sum over $n$ in Eq.
\ref{evolution1} can be replaced by an integration over $\epsilon$, the
statistical blocking parameter, namely
\begin{eqnarray}
   \sum_n \to \int d\epsilon M(\epsilon)
\label{n-to-eps}
\end{eqnarray}
with $M(\epsilon)$ the integration measure of the transformation.
Eq. \ref{evolution1} becomes
\begin{eqnarray}
        \braket{\phi_0,t=+\infty}{\phi_0,t=-\infty} &=&
           \int d\epsilon M(\epsilon)\sum_{\xi\xi'} \widetilde C^\epsilon_\xi
         \widetilde C^{\epsilon *}_{\xi'}
         \braket{n\xi',t=+\infty}{n\xi,t=-\infty}\nonumber\\
         &=& \int d\epsilon  \int \prod_i D[f_i]
                   e^{iS_{eff}[f,\mu,\epsilon] +
                    {1\over 2} \overline\eta S_F[f] \eta
                     + i\int d^4x \left (\sum_k J_k f_k
                      - V_0(\mu,\epsilon) \right )}
\label{General_W_3}
\end{eqnarray}
with $exp[-i\int d^4x V_0(\mu,\epsilon)]$ the leading piece of the
measure for the $\epsilon$ integration in the thermodynamic
limit. When the volume $L^3$ of the system becomes increasingly large,
the integrand of the $\epsilon$ integration becomes increasingly sharp
at the extrema positions of the argument of the exponential in the
above equation. This is due to the fact that $\epsilon$ couples to
macroscopic variables that are proportional to the volume; it has no
quantum fluctuation in the thermodynamic limit. Due to this reason,
the detailed form of the measure $M(\epsilon)$ but its leading piece
in Eq. \ref{n-to-eps} is irrelevant in the thermodynamic limit. So
the weighted sum
\begin{eqnarray}
       \sum_{\{\phi_0\}} {\cal W}[\phi_0,\phi_0]
         \braket{\phi_0,t=+\infty}{\phi_0,t=-\infty} &=& \int \prod_i
         D[f_i] e^{iS_{eff}[f,\mu,\epsilon] + {1\over 2} \overline\eta
         S_F[f] \eta + i\int d^4x \left (\sum_k J_k f_k -
         V_0(\mu,\epsilon) \right )}
\label{General_W_4}
\end{eqnarray}
with $\epsilon$ taking the value of one of the extrema of the argument
of the exponential. The conventional method, which may turns out to be
not sufficient in symmetry breaking phases, corresponds to a special
case, namely, the $\epsilon=0$ one. The relevance of introducing
$\epsilon$ will be discussed in the following sections.
It can be seen that a possible finite $\epsilon$ in the final result
is perturbatively non-reachable.

\subsubsection{The determination of $V_0(\mu,\epsilon)$}

Eq. \ref{General_W_4} tells us that the generating functional of an 
interacting fermionic system should be written as
\begin{eqnarray}
     e^{W[J,\overline\eta,\eta,\mu,\epsilon]} &=&
       \int \prod_i D[f_i]
                   e^{iS_{eff}[f,\mu,\epsilon] +
                    {1\over 2} \overline\eta S_F[f] \eta
                     + i\int d^4x \left (\sum_k J_k f_k
                     - V_0(\mu,\epsilon) \right )}.
\label{General_W_5}
\end{eqnarray}
It is normalized by the condition
\begin{eqnarray}
     W[0,0,0,0,0] &=& 0.
\label{W-normal}
\end{eqnarray}
For a stationary situation, $\widetilde S_{eff}$ corresponding to Eq.
\ref{tS_eff} can be expressed (see also Eq.
\ref{Seff2}) as
\begin{eqnarray}
    \widetilde S_{eff}[f,\mu,\epsilon] 
     &=& -i{T\over 2}\left ( \sumint {dp^0\over 2\pi}
    \ln{\lambda_{p^0,\xi}[f,\mu]\over
     \lambda_{p^0,\xi}[0,\mu]} + \sum_{\xi} \int_{\cal C} {dp^0\over 2\pi}
    \ln\lambda_{p^0,\xi}[0,\mu] - \sum_{\xi} \int_{{\cal C}_0} {dp^0\over 2\pi}
    \ln\lambda_{p^0,\xi}[0,0] \right )\nonumber\\ &&+
    \int d^4 x [{\cal L}_B[f]-V_0(\mu,\epsilon)],
\label{Seff_FD2}
\end{eqnarray}
where ${\cal C}$ represents the $p^0$ integration contour shown in Fig.
\ref{Fig:ConFDB0} and ${\cal C}_0$ represents the $p^0$ integration contour
given by Fig. \ref{Fig:ConCon1}. $V_0(\mu,\epsilon)$ satisfies
$V_0(\mu,0)= \int d^4 x \mu_\alpha \overline j^\alpha$ so that it
agrees with Eq. \ref{tS_eff} in this special case. From
Appendix \ref{app:FF}, it is shown that the left hand side of the
following equation is finite for an uniform system, namely, 
\begin{eqnarray}
 -i{T\over 2}\left (\sum_{\xi} \int_{\cal C} {dp^0\over 2\pi}
    \ln\lambda_{p^0,\xi}[0,\mu]
    \right. &-& \left . \sum_{\xi} \int_{{\cal C}_0} {dp^0\over 2\pi}
    \ln\lambda_{p^0,\xi}[0,0] \right ) = \nonumber \\ &&
         \int d^4 x \left [
         \left (\mu_+ \overline\rho_{(+)} + \mu_- \overline\rho_{(-)}
                            - \mu \overline\rho \right ) - \left (
                          \overline e_{(+)} + \overline e_{(-)} -
                          \overline e \right ) \right ].
\end{eqnarray}

 The {\em basic assumption} of the theory is then the following 
{\em choice} for $V_0(\mu,\epsilon)$, namely,
\begin{eqnarray}
     V_0(\mu,\epsilon) &=& \int d^4 x
         \left (\mu_+ \overline \rho_{(+)} + \mu_- \overline\rho_{(-)}
                             - \mu \overline \rho \right )
\label{V_0_Form}
\end{eqnarray}
which completely specifies Eq. \ref{General_W_4}. For an uniform
system, the corresponding effective potential in a  Hartree--Fock
approximation to be minimized is
\begin{eqnarray}
   V_{eff} &=&
\lim_{V_3\to\infty}
i{1\over 2V_3} \sumint {dp^0\over 2\pi}
    \ln{\lambda_{p^0,\xi}[f;\mu]\over
\lambda_{p^0,\xi}[0,\mu]}
     +{n_f  n_c\over 4\pi^2} \left (\mu^4 + 2\epsilon^4 + 12 \mu^2\epsilon^2
   \right )
     -  {\cal L}_B[f]
\label{Veff-FDB}
\end{eqnarray}
with ${\cal C}$ denoting the $p^0$ integration contour chosen.

In order to preserve the causal structure of the original Minkowski
$p^0$ integration contour, the Euclidean effective action is obtained
by distorting the Minkowski contour given in Fig. \ref{Fig:ConFDB0} to 
the one labeled ``II'' in Fig. \ref{Fig:ConFDB1}.

\subsection{More on the primary statistical gauge field}
\label{subsec:more}
\subsubsection{Statistical gauge invariance, physical states
             and conservation of fermion number}

   In the process of introducing the primary statistical gauge field
$\mu^\alpha$, the original global
$U(1)$ symmetry corresponding to the fermion number conservation is replaced,
in a certain sense, by a local symmetry. This local symmetry
originates from
the fact that the eigenvalues $\lambda_{p^0,\xi}[f,\mu]$,
which satisfies the eigenequation
\begin{eqnarray}
       \gamma^0 iS_F^{-1}[f,\mu] \Psi_\lambda =
       \lambda[f,\mu] \Psi_\lambda,
\label{EigenEq3}
\end{eqnarray}
is invariant under the following gauge transformation
\begin{eqnarray}
       \Psi_\lambda(x) &\to & e^{i\phi(x) O_3}
       \Psi_\lambda(x)\label{G-tran1}\\
       \mu^\alpha(x) &\to & \mu^\alpha(x) - \partial^\alpha\phi(x)
       \label{G-tran2}
\end{eqnarray}
with $\phi(x)$ an arbitrary function of the spacetime that decreases
to zero sufficiently fast at the spacetime infinity. The
introduction of a local field $\mu^\alpha(x)$ is expected to
introduce infinite extra degrees of freedom, which should be
eliminated in certain way. Albeit the full effective action given by
Eq. \ref{Seff_FD2}, which depends on $\mu^\alpha\mu_\alpha\ge 0$,
is not invariant under the gauge transformations given by
Eqs. \ref{G-tran1} and \ref{G-tran2},
the primary statistical gauge invariance of the quantum fluctuation or
the connected part of $\widetilde S_{eff}$ requires further investigation.

Let us find the connection of the primary statistical gauge
invariance to the conservation of fermion number by
quantize the system governed by the Lagrangian density
\begin{eqnarray}
     {\cal L}' &=& {\cal L} + \mu^\alpha j_\alpha
\label{Lag-prim}
\end{eqnarray}
in Eq. \ref{General_W_5}. Here
${\cal L}$ is the original Lagrangian density before introducing
$\mu^\alpha$ and $j_\alpha$ is the fermion number current density
given by Eq. \ref{fermion-curr}.
The fermion field $\Psi$ and the boson fields $\{f_i\}$
are quantized as usual \cite{TFTQFT}; they shall not be repeated here.

What is needed to be found here is the conjugate variable
$\pi_{u}^\alpha$ of $\mu^\alpha$.
For that purpose,
Eq. \ref{Lag-prim} can be treated as the Hamiltonian
density, namely
\begin{eqnarray}
   {\cal H}_{(\mu)} &=& {\cal L}'.
\label{Hamilt}
\end{eqnarray}
 Before the quantization, it follows
from the Hamiltonian dynamics that
\begin{eqnarray}
\partial_0{\pi}_{ui} &=& -{\partial {\cal H}_{(\mu)}\over\partial \mu_i} = j_i
,\label{qu-dot}\\
\partial_0{\mu}_{i} &=& {\partial {\cal H}_{(\mu)}\over\partial \pi_{ui}} = 0
,\label{uu-dot}\\
\end{eqnarray}
where $i,j=1,2,3$ labels the spatial components of 4-vectors.

The quantization of $\mu_i$ is then implemented by the Dirac
quantization condition
\begin{eqnarray}
   \left [\widehat\pi_{ui}({\bf x},t), \widehat\mu_j({\bf x}',t) \right ]
      &=& - i \delta^{(3)}({\bf x}-{\bf x}')\delta_{ij}.
\label{mui-quantization}
\end{eqnarray}
The {\em statistical ``electric field''} $\widehat\pi_{ui}$ commutes
with all elementary fields in the Lagrangian density but the one
listed above at equal-time. Note that all quantities with a hat
``$\wedge$'' on top denote operators in the following.

After the quantization, the set of gauge transformation, in which $\phi(x)$
is independent of time, is represented by
\begin{eqnarray}
    \widehat\Psi &\to& U[\phi] \widehat \Psi U^\dagger[\phi] =
    e^{i\phi O_3}\widehat \Psi \label{G-tran3}\\ \widehat \mu_\alpha
    &\to& U[\phi] \widehat \mu_\alpha U^\dagger[\phi] = \widehat
    \mu_\alpha - \partial_\alpha\phi \label{G-tran4}
\end{eqnarray}
with
\begin{eqnarray}
     U[\phi] &=& e^{-i\int d^3 x
         \left ( \widehat \rho + \nabla\cdot\widehat{\bf
         \pi}_{u} \right )\phi},
\label{Us-operator}
\end{eqnarray}
where the space integration is at any specific time at which a
transformation of an operator is considered.
The operator algebra between dynamical fields is then represented in
a Hilbert space. Since the quantity $\mu^\alpha$ is
introduced into the original theory, it is expected that this Hilbert
space contains states that are not physical or are redundant.
The physical states are selected within the full Hilbert space
by requiring that they satisfy
\begin{eqnarray}
     \bra{\mbox{Phys}'}U[\phi]\ket{\mbox{Phys}}
         &=& Z e^{-i\Phi}
\label{C-consv1}
\end{eqnarray}
under the time independent gauge transformation discussed above, with
the common phase factor $\Phi$ restricted to those functions that are
independent of time (it is explained in the following).  Taking the
time derivative of Eq. \ref{C-consv1}, and using dynamical equation
Eq. \ref{qu-dot}, one obtains
\begin{eqnarray}
     \bra{\mbox{Phys}'}
        \left ( \partial_0{\widehat\rho} + \nabla\cdot \widehat {\bf
         j}\right ) \ket{\mbox{Phys}}
       &=& \bra{\mbox{Phys}'}\partial^\mu \widehat j_\mu \ket{\mbox{Phys}} = 0
\label{C-consv3}
\end{eqnarray}
which is the conservation of fermion number current in physical processes.

The time independent and spatially localized gauge transformation considered
is non-trivial one. It selects
amongst those states in the extended Hilbert space the physical ones.
This can be understood if one consider the commutation relation
between $\widehat \rho + \nabla \cdot \widehat\pi_{u}$ and the Hamiltonian of
the system.
\begin{eqnarray}
  \left [\widehat \rho + \nabla \cdot \widehat\pi_{u},\widehat H \right ]
        &=& i{d\over dt} \left (\widehat \rho + \nabla \cdot \widehat\pi_{u}
      \right ) = \partial_\mu \widehat j^\mu
\label{H-Q-comm}
\end{eqnarray}
with $\widehat H$ the total Hamiltonian of the system.  It is zero due
to the conservation of fermion number current. It means that $\widehat
\rho + \nabla \cdot \widehat\pi_{u}$ is independent of time when the
matrix elements between physical states are taken.  The states in the
extended Hilbert space can be divided into subspaces labeled by a
complex (time independent) function of the spatial coordinates
according to the matrix elements of $ \widehat \rho + \nabla
\cdot \widehat\pi_{u}$ between themselves. For those eigenstates of the
Hamiltonian of the system, it can be written
as\footnote{It is the quantum version of the ``classical''     
constraint equation $\rho + \nabla \cdot \pi_{u}
=\varsigma$.}
\begin{eqnarray}
     \bra{\varphi^i_\varsigma}
       \left (\widehat \rho + \nabla \cdot \widehat\pi_{u}  \right )
      \ket{\varphi^j_\varsigma}
      &=& \delta_{E_i E_j} N_{ij} \varsigma
\label{Q-eigenstate}
\end{eqnarray}
with $\varsigma$ the space dependent complex function,
$\ket{\varphi^k_\varsigma}$ a state in the physical space that has
energy $E_k$, $\delta_{EE'}$ taking zero or unity value if $E\ne E'$
or $E'=E$ (assuming that $E$ is discrete before thermodynamic limit
is taken) and $N_{ij}$ independent of spacetime
coordinates\footnote{Eq. \ref{H-Q-comm} may not be restrictive
enough. Somewhat more restrictive constraint can be suggested. It
consists of decomposing $\widehat \rho + \nabla \cdot \widehat\pi_{u}$
into superposition of
holomorphic $(\widehat \rho + \nabla \cdot \widehat\pi_{u})_{(-)}$
and antiholomorphic $(\widehat \rho + \nabla \cdot
\widehat\pi_{u})_{(+)}$components \cite{JKG}.  The physical states are
those ones that are eigenstates of $(\widehat \rho + \nabla \cdot
\widehat\pi_{u})_{(-)}$ with a common eigenvalue $\varsigma$. }. Eqs.
\ref{H-Q-comm} and \ref{Q-eigenstate} mean that the physical states
are the ones that have vanishing matrix elements on the commutator of
the Hamiltonian and $\widehat \rho + \nabla \cdot \widehat\pi_{u}$.
Therefore they are expressible by a superselection sector in the
extended Hilbert space defined and labeled by a complex function
$\varsigma$ of spatial coordinates.  Such a definition of the physical
states for the primary statistical gauge theory is less restrictive
than the one used in dynamical gauge theory like QED \cite{GSBK} where
due to the existence of the dynamical part for the gauge fields at the
tree level, the physical states are restricted to the subspace with
$\varsigma\equiv 0$ only. In fact, for the $\beta$ and $\omega$ phases
discussed in the following, in which the $U(1)$ symmetry corresponding
to the fermion number conservation is spontaneously broken down,
$\varsigma$ can not be zero due to the fact that before taking into
account of the dynamical gauge fields (that of the photon), the
massless Goldstone boson corresponding to the spontaneous symmetry
breaking has to be considered as a physical excitation. But if
$\varsigma$ is chosen to be zero, the massless Goldstone boson belongs
to unphysical states \cite{Strocchi}. Such a situation actually opens
up the possibility for the spontaneous CP violation to be discussed in
the following. The choices made for the physical states is a
constraint invariant under time evolution due to
Eq. \ref{H-Q-comm}. It shows that definition for physical states
remains true at all times and no transition to unphysical states and
between the superselection sectors is possible in physical processes.

\subsubsection{The statistical gauge degrees of freedom and the
  question of long range order}

   The primary statistical gauge field $\mu^\alpha$ is non-dynamical
at the tree level. At the quantum level a dynamics for $\mu^\alpha$ is
generated due to the fermion quantum fluctuations. The relevant
effective action for the primary  statistical gauge field can be obtained from
Eq. \ref{Seff_FD2} by a ``Legendre transformation'' to a form without
the contribution of $V_0$, namely
\begin{eqnarray}
S_{eff}[f,\mu,\epsilon] 
     &=& -i{T\over 2}\left ( \sumint {dp^0\over 2\pi}
    \ln{\lambda_{p^0,\xi}[f,\mu]\over \lambda_{p^0,\xi}[0,\mu]} +
    \sum_{\xi} \int_{\cal C} {dp^0\over 2\pi}
    \ln\lambda_{p^0,\xi}[0,\mu] - \sum_{\xi} \int_{{\cal C}_0}
    {dp^0\over 2\pi} \ln\lambda_{p^0,\xi}[0,0] \right )\nonumber\\ &&+
    \int d^4 x {\cal L}_B[f],
\label{Seff_mu}
\end{eqnarray}
which is a canonic functional of $\mu^\alpha$.
The quadratic term for slow varying
$\mu^\alpha$ (in spacetime or at long distances) 
generated from the fermion determinant is
of the form
\begin{eqnarray}
     S^{(\mu)}_{eff}
      &=& {1\over 2}\int d^4x d^4x' \mu'_\alpha(x) \pi^{\alpha\beta}(x,x')
      \mu'_\beta(x') + {n_f n_c\over \pi^2} \int d^4 x
             \epsilon^2 {\mu'}^2
,\label{Seff-mu1}
\end{eqnarray}
where $\mu'_\alpha = \mu_\alpha - \overline\mu_\alpha$ with
$\overline\mu_\alpha$ shifted $\mu_\alpha$ and the last term is from the
corresponding one in Eq. \ref{Seff_mu}. The first term is generated
from the fermion determinant. 
If the electromagnetic interaction between the fermions are not
considered, $\pi^{\alpha\beta}(x,x')$ is given by
\begin{eqnarray}
    \pi_0^{\alpha\beta}(x,x') &=& i\bra{0} T j^\alpha(x) j^\beta(x')
    \ket{0} \nonumber \\
     &=&  Z^{(\mu)}(\partial_x^2 g^{\alpha\beta} -
    \partial_x^\alpha\partial_x^\beta ) \delta(x-x') + i\bra{0}
    j^\alpha(x)\ket{0} \bra{0} j^\beta(x') \ket{0}
\label{normal-pi-mn}
\end{eqnarray}
in the normal phase; and,
\begin{eqnarray}
    \pi_0^{\alpha\beta}(x,x') &=&  i\bra{0} T j^\alpha(x) j^\beta(x')
    \ket{0} \nonumber \\
     &=& \Pi^{(\mu)}\left (g^{\alpha\beta}
    -{\partial_x^\alpha\partial_x^\beta\over\partial^2} \right )
    \delta(x-x')+ i\bra{0}
    j^\alpha(x)\ket{0} \bra{0} j^\beta(x') \ket{0},
\label{sup-pi-mn}
\end{eqnarray}
in the phase where the $U(1)$ symmetry corresponding to the fermion
number conservation is spontaneously broken down since there exists a
massless pole in the matrix element of $j^\alpha$ (see
Ref. \cite{Ying11} for a more detailed discussion). Here $Z^{(\mu)}$ and
$\Pi^{(\mu)}$ are functions of $x$ and $x'$.

In the normal phase, the effective action for slow varying
$\mu'_\alpha$ is
\begin{eqnarray}
      S^{(\mu)}_{eff} = \int d^4 x \left [ -{Z^{(\mu)}\over 4} f_{\mu\nu}
      f^{\mu\nu} + {1\over 2} \left( i\bra{0}j^\alpha\ket{0}\bra{0}j^\beta\ket{0}
       + 2 g^{\alpha\beta} {n_f n_c\over \pi^2} \epsilon^2 \right )
       \mu'_\alpha \mu'_\beta \right ]
\label{Seff-mu2}
\end{eqnarray}
with $f^{\alpha\beta} = \partial^\alpha
{\mu'}^\beta-\partial^\beta{\mu'}^\alpha$, since the eigenvalues 
$\lambda_{p^0,\xi}$ are invariant under the gauge
transformation given by Eqs. \ref{G-tran1} and \ref{G-tran2}. In the
phase where $U(1)$ is
spontaneously broken down and before considering electromagnetic interaction,
\begin{eqnarray}
      S^{(\mu)}_{eff} &=& {1\over 2}\int d^4 x \left \{ \left [
             {i}\bra{0}j^\alpha\ket{0}\bra{0}j^\beta\ket{0}
       + g^{\alpha\beta} \left ({\Pi^{(\mu)}} + 2{n_f n_c\over \pi^2}
       \epsilon^2 \right )
       \right ] \mu'_\alpha
      \mu'_\beta  + \ldots\right \}.
\label{Seff-mu3}
\end{eqnarray}
In both of the situations with slow varying $\mu'_\alpha$, $Z^{(\mu)}$ and 
$\Pi^{(\mu)}$ are approximately constants.

If the electromagnetic interaction is considered, then there is a
$f_{\alpha\beta} f^{\alpha\beta}$ term in Eq.  \ref{Seff-mu3} and the
$\Pi^{(\mu)}$ term is absent, even in a spontaneous $U(1)$ symmetry
breaking phase.  This is because when the electromagnetic interaction
between the fermions are considered, $\pi^{\alpha\beta}$ can be
decomposed into a connected and disconnected part or
$\pi^{(c)}_{\mu\nu} + i\bra{0} j_\mu \ket{0} \bra{0}j_\nu \ket{0} $
with the connected part given by
\begin{eqnarray}
    \pi^{(c)}_{\mu\nu} &=& \pi^{(c)}_{0\mu\nu} + \left (
             \pi^{(c)}_0 G_0 \pi^{(c)}_0\right )_{\mu\nu}+ 
     \left ( \pi^{(c)}_0 G_0 \pi^{(c)}_0
              G_0 \pi^{(c)}_0\right )_{\mu\nu} +
     \ldots
\end{eqnarray}
and
\begin{eqnarray}
      \pi^{(c)}_{0\mu\nu} &=& (q^2 g_{\mu\nu} - q_\mu q_\nu) \pi_0^{(c)}
      \nonumber\\
      G_0^{\mu\nu} &=& \left ( - g^{\mu\nu} + {q^\mu q^\nu\over q^2}
                     \right ) {i\over q^2},
\end{eqnarray}
where $G^{\mu\nu}_0$ is the propagator of the bare photon.
$\pi^{(c)}_{\mu\nu} \to 0$ in the $q^\mu\to 0$ limit even when
$\pi_{0\mu\nu}^{(c)}$ contains a massless pole (in $q^2$).  Whatever
the case, it can be easily seen that the full two point proper vertex
for $\mu'_\alpha$ is non-vanishing in the $q^\mu\to 0$ limit if either
$\epsilon\ne 0$ or $\bra{0}j_\mu\ket{0}\ne 0$ or both in the vacuum
state of the system. Therefore a long range order for $\mu'_0$ is
possible only if both $\epsilon=0$ and $\mu=0$ in the vacuum. The
spatial component $\mbox{\boldmath $\mu$}'$ of $\mu'_\alpha$ is short
ranged in the phase where $\epsilon\ne 0$ and $\mu_0=0$. It is however
long ranged in the phase where $\epsilon=0$.

\subsubsection{Topological configurations}

  The discussion given above shows that after the introduction of a
primary statistical gauge field $\mu_\alpha$, the representation Hilbert space
of the operator algebra is extended. Such an extended Hilbert space
includes not only the physical states, but also non-physical ones. The
physical states are the ones that satisfy Eq.  \ref{Q-eigenstate},
which is expected to be satisfied in all consistent computations of
physical observable where fermion number current conservation is
preserved.

   The extension of the representing Hilbert space gives us additional
leverage to project out the collective excitations and configurations
not easily discernible in the conventional approach.

    Due to the statistical gauge invariance under the local
transformation given by Eqs. \ref{G-tran1} and \ref{G-tran2}, the
physically non-trivial configurations of the system depend, in the
path integration sense, only on the flux density or statistical
``magnetic field'' \mbox{\boldmath $b$} defined as
\begin{eqnarray}
    \mbox{\boldmath $b$} &=& \nabla\times \mbox{\boldmath$\mu$}.
\label{b-def}
\end{eqnarray}
Therefore the general form of the generating functional Eq. \ref{General_W_5}
can be written more precisely by imposing certain gauge fixing
condition or as
\begin{eqnarray}
     e^{W[J,\overline\eta,\eta,\mu,\epsilon]} &=&
       \int D[\mbox{\boldmath $b$}] J(\mbox{\boldmath $b$})\prod_i D[f_i]
                   e^{iS_{eff}[f,\mu,\epsilon] +
                    {1\over 2} \overline\eta S_F[f] \eta
                     + i\int d^4x \left (\sum_k J_k f_k
                     - V_0(\mu,\epsilon) \right )},
\label{General_W_6}
\end{eqnarray}
where $J(\mbox{\boldmath $b$})$ is the integration measure. Eq.
\ref{General_W_6} is equivalent to Eq.  \ref{General_W_5}. It is
however useful for us to find out collective stationary configurations
that are not easily found in a common approach to the effective
action.

For a configuration in which \mbox{\boldmath $b$} is non-vanishing
only in a localized region, quantization of flux appear, namely
\begin{eqnarray}
    \int_{\Sigma} d{\mbox{\boldmath $S$}}\cdot {\mbox{\boldmath $b$}}
     &=& \oint_{\partial \Sigma}
    d{\mbox{\boldmath $l$}}\cdot{\mbox{\boldmath $\mu$}}
    = 2n\pi, \hspace{0.6cm}(n=0,\pm
    1,\pm 2,\ldots)
,\label{flux-quant}
\end{eqnarray}
where $\Sigma$ is the surface that contains the \mbox{\boldmath $b$}
and line integration is around the edge $\partial\Sigma$ of
$\Sigma$. The quantization results, as it is well known, by imposing
the uniqueness condition on the eigenfunctions $\Psi_\lambda[f,\mu]$.

\subsubsection{A new macroscopic parameter and long range order}

Using the primary statistical gauge field $\mu^\alpha$, a new
macroscopic parameter that characterizes the vacuum state of the
system can be introduced. It is defined as
\begin{eqnarray}
   \widehat O_\Sigma &=& e^{
       i\oint \displaystyle d\mbox{\boldmath $l$}\cdot
       \mbox{\boldmath
       $\mu$} },
\end{eqnarray}
where $\Sigma$ is a large 2-dimensional surface area in space at
certain time and the line integration $\oint$ is along the edge of the
area $\Sigma$.

 It is known \cite{JBKogut} that the vacuum expectation value of
$\widehat O_\Sigma$ provides another one of the macroscopic parameters
for a more detailed characterization of the phase structure of the
system. For example, in a disordered system in which the correlation
between the complex phase of the fermions at different space points
becomes short ranged, condensation of vortices or monopoles of the
type of configurations with non-vanishing $n$ in Eq. \ref{flux-quant}
can derive a Kosterlitz--Thouless \cite{KT-phase} type of phase
transition.

\section{Two Models for Strong Interaction
          and Their Vacuum Phase Structure}
\label{sec:Models}

 The fundamental theory for the strong interaction is considered to be
QCD with its Lagrangian density given in Appendix \ref{app:QCD}.  The
current available method of studying QCD from first principle is the
lattice QCD simulation. Albeit great progresses are made, the lattice
computation are limited by the small lattice sizes and by the
limitation in the computer power. Model approaches, which are simpler
than the full QCD calculation and has large overlap with it in the low
and intermediate energy regions can be and have been used to obtain
much of the physical pictures that are supposed to happen in the
system governed by QCD Lagrangian density. Since the mass of the light
quarks have values much smaller than the typical scale of the hadronic
spectrum of order 1 GeV, certain subset of the behaviors of the light
quark system can be simulated by an interacting massless fermion
systems that possesses the basic symmetries of the QCD Lagrangian
density. It is expected that we can learn some of the important
possible behaviors of the light quark system by using models due to
our (relatively) increased theoretical analytic power. The spontaneous
chiral $SU(2)_L\times SU(2)_R$ symmetry breaking down in hadronic
systems was historically discussed before the birth of the conception
of quark and QCD Lagrangian density. This phenomenon is only latter
justified by the QCD (lattice) calculation.

I discuss in this paper the physical properties of strong interaction
vacuum related to the fluctuation of baryon number using two model
Lagrangian densities that possess the chiral $SU(2)_L\times SU(2)_R$
symmetry of massless QCD and has 3 colors ($n_c=3$).  For a full
investigation of the possible phases of the vacuum of a relativistic
massless fermion system, these models are so chosen that they allow
not only the quark--antiquark condensation that is widely discussed in
the literature but also the rarely studied phases that are induced by
a condensation of quark--quark (or antiquark--antiquark) pairs. One of
these possibilities is studied in detail in Ref. \cite{Ying1,Ying11}.

    In order to simplify our discussion, I consider two model
Lagrangian densities that are half bosonized. Both of them can be
viewed as been the descendents of two four quark interaction models
after introducing the auxiliary fields in a Fierz invariant way
\cite{Ying11}.

This section, which mainly serves the purpose of introducing the
models, contains the determination of their vacuum phase structure
using the conventional approaches.  A more detailed study using the
refined method developed in sections \ref{sec:Instab} and
\ref{sec:FDth} is given in the next section.

\subsection{Model I and the $\omega$ phase}

The first model is defined by the following Lagrangian density
\begin{eqnarray}
 {\cal L}_1 & = & {1\over 2} \overline \Psi\left [i{\rlap\slash\partial}-\sigma-
             i\vec{\pi}\cdot \vec{\tau}\gamma^5 O_3-\gamma^5 {\cal
             A}_c\chi^c O_{(+)}-\gamma^5 {\cal A}^c\overline\chi_c O_{(-)}
             \right ]\Psi   -{1\over 4 G_0} (\sigma^2 +
             \vec{\pi}^2) + {1\over 2 G_{3'}} \overline\chi_c \chi^c
,\label{Model-L-1}
\end{eqnarray}
where $\sigma$, $\vec{\pi}$, $\overline\chi_c$ and $\chi^c$ are
auxiliary fields with $(\chi^c)^{\dagger} = - \overline\chi_c$ and
$G_0$, $G_{3'}$ are coupling constants of the model.  ${\cal A}_c$ and
${\cal A}^c$ $(c=1,2,3)$ act on the color space of the quark; they are
\begin{eqnarray}
{\cal A}_{c_1c_2}^c = -\epsilon^{cc_1c_2}\hskip 0.5 in &&
{\cal A}_{c,c_1c_2} = \epsilon^{cc_1c_2}\label{Amtr}
\end{eqnarray}
with $\epsilon^{abc}$ ($a,b = 1,2,3$) the total antisymmetric
Levi--Civit\'a tensor. Here the quark spinor is represented by the
8-dimensional Dirac spinor and $O_{(\pm)}$ are raising and lowering
operators respectively in the upper and lowering 4 components of
$\Psi$.

The effective action 
is given by Eq. \ref{Seff2} with the auxiliary fields independent of
spacetime, the effective potential has the following form
\begin{eqnarray}
    V_{eff} &=& -\lim_{L,T\to\infty} {1\over L^3 T} S_{eff}\nonumber\\
            &=& -{i\over 2} \lim_{L,T\to\infty} {1\over L^3 T}
                 \sum_{\lambda_n} \ln{\lambda_n\over\lambda_n^{(0)}}
                  + {1\over 4 G_0} \sigma^2 - {1\over 2 G_{3'}}
                  \overline\chi_c\chi^c
,\label{V-eff-2}
\end{eqnarray}
where $\lambda_n$ and $\lambda_n^{(0)}$ correspond to the eigenvalues
of the two Hermitian operators defined in Eq. \ref{EigenEq} with and
without the auxiliary fields shifted respectively.  Since the
auxiliary fields do not depend on spacetime, the eigenvalues
$\lambda_n$ and $\lambda_n^{(0)}$ can be labeled by the 4-momentum of
the corresponding eigenstates $\Psi_{\lambda_n}$. The result is
\begin{eqnarray}
    V_{eff} &=& {i\over 2} \int {d^4p\over (2\pi)^4} \left (
                 8 \sum_{i=1}^4
                 \ln{\lambda_i(p)\over\lambda_i^{(0)}(p)}
                + 4 \sum_{i=1}^4
                 \ln{{\lambda'}_i(p)\over{\lambda'}_i^{(0)}(p)}
                \right ) + {1\over 4 G_0} \sigma^2 - {1\over 2 G_{3'}}
                \overline \chi_c \chi^c\label{V-eff-2-1},
\end{eqnarray}
where $\lambda_i(p)$, ${\lambda'}_i(p)$ are eigenvalues of states
with color different, the same as $\chi^c$ respectively and factors 8,
4 correspond to their degeneracy. It has the following explicit form
\begin{eqnarray}
    V_{eff} &=& 4i\int {d^4p\over (2\pi)^4} \ln\left [\left (
            1- {\sigma^2+\chi^2\over p^2} \right )^2-{\sigma^2\over
              p^2}
              \left ( 1-{\sigma^2-\chi^2\over p^2} \right )^2\right ]
               + {1\over 4 G_0}\sigma^2 + {1\over 2G_{3'}} \chi^2
,\label{Veff-2-2}
\end{eqnarray}
where $\chi^2 \equiv -\overline\chi_c\chi^c$.

In the Minkowski spacetime formulation, the contour for the $p^0$
integration is shown in Fig. \ref{Fig:ConCon1}. One of the most
commonly used ones, which is called quasiparticle path, is shown in
Fig. \ref{Fig:ConCl2}. It can be shown that the resulting effective
potential is the change of the energy density of the quasiparticles,
which is obtained from summing over the quasiparticle energies
determined by the poles of $S_F[f]$, relative to that of the bare
particle in the truncated Dirac sea. It does not has a covariant form
due to the fact that a non-covariant cutoff in the 3-momentum ${\bf
p}$ space has to be introduced. The Euclidean path shown in Fig.
\ref{Fig:ConCl2} can be used to obtain a covariant expression for the
effective action. The Euclidean covariant cutoff scheme does not result
in deriving physically undesirable results in other covariant
approaches \cite{Hatsuda}.  As discussed in section 2.2, it can also
be used to include quantum effects (see Appendix \ref{app:thedark})
that are beyond the quasiparticle approximation. The differences
between these two paths are however not very important for the purpose
of this subsection. Since different kind of cutoffs are used in these
two approaches, a comparison between them is difficult. Nevertheless,
I shall adopt the Euclidean path shown in Fig. \ref{Fig:ConCl2}. An
Euclidean path is however important for discussions to be
followed. The resulting expression is
\begin{eqnarray}
    V_{eff}(\sigma,\chi) &=& -4\int^\Lambda {d^4p\over (2\pi)^4}
    \ln\left [\left (
            1+{\sigma^2+\chi^2\over p^2} \right )^2+{\sigma^2\over
              p^2}
              \left ( 1+{\sigma^2-\chi^2\over p^2} \right )^2\right ]
               + {1\over 4 G_0}\sigma^2 + {1\over 2G_{3'}} \chi^2
,\label{Veff-2-3}
\end{eqnarray}
where $\Lambda$ is the Euclidean cutoff introduced to define the
model. A numerical evaluation shows that the minima of
$V_{eff}(\sigma,\chi)$ is located on either the $\sigma$ axis
($\chi=0$) or the $\chi$ axis ($\sigma=0$). Explicit expression for
$V_{eff}(\sigma,0)$ and $V_{eff}(0,\chi)$ are found to be
\begin{eqnarray}
     v_{eff}(\sigma,0) &=& 3 f({\sigma^2\over\Lambda^2}) + {1\over
       16\pi\alpha_0} {\sigma^2\over\Lambda^2}\label{Veff10}\\
     v_{eff}(0,\chi)   &=& 2 f({\chi^2\over\Lambda^2}) + {1\over
       16\pi\alpha_{3'}} {\chi^2\over\Lambda^2}
,\label{Veff01}
\end{eqnarray}
where the dimensionless effective potential $v_{eff}$ is defined by
$V_{eff}\equiv \Lambda^4 v_{eff}$,
$\alpha_{0} = G_0\Lambda^2/4\pi$ and $\alpha_{3'} =
G_{3'}\Lambda^2/8\pi$ with 
\begin{eqnarray}
     f(x) &=& {1\over 8\pi^2}\left [-x + \ln\left (1+{1\over x}\right)
     x^2 - \ln(1+x) \right ]. \label{f-func}
\end{eqnarray}
The values of $\sigma^2$ and $\chi^2$ at the minima of
Eqs. \ref{Veff10} and \ref{Veff01} determine the vacuum of the system
in the one loop Hartree--Fock approximation for the fermions
(i.e. Eq. \ref{1-loop-gamma} without the second bosonic one loop
term). The phase structure of the model is shown in the
$\alpha_0$--$\alpha_{3'}$ plane in Fig. \ref{Fig:bound1}. Three kinds
of phases for the vacuum are possible. The first phase, which is
called the $O$ phase, is the bare vacuum. The second phase, or the
$\alpha$ phase, has non-vanishing vacuum expectation value of
$\overline\Psi\Psi$; the chiral $SU(2)_L\times SU(2)_R$ symmetry is
spontaneously broken down to a $SU(2)_V$ flavor symmetry. The third
phase, called the {\em $\omega$ phase} of the vacuum, has
non-vanishing diquark and antidiquark condensation characterized by a
non-vanishing $\chi^2$; chiral symmetry is unbroken in this phase.

The phase transitions across the boundary between the $O$ and the
$\alpha$ phases ($\alpha_{0}=\pi/12$ and $\alpha_{3'}<\pi/8$) and the
one between the $O$ and the $\omega$ phases ($\alpha_0<\pi/12$ and
$\alpha_{3'} = \pi/8$) are of second order. The phase transition
between the $\alpha$ and the $\omega$ phases ($\alpha_0>\pi/12$ and
$\alpha_{3'} > \pi/8$) is of first order. The Meissner effects for the
electromagnetic field are expected in the $\omega$ phase. The basic
physics of it is discussed in more detail in
Refs. \cite{Ying1,Ying11,GDHp} for model II of the following. I shall
relegate such a discussion for this model to other work.

The Minkowski
propagator for the quarks in the $\omega$ phase can be found by an
inversion of the operator $S_F^{-1}[f]$ in the action, namely.
\begin{eqnarray}
     S_F &=& i\left [ i{\rlap\slash\partial}-\gamma^5 {\cal
             A}_c\chi^c O_{(+)}-\gamma^5 {\cal A}^c\overline\chi_c O_{(-)}
                  \right ]^{-1}.
\label{Prop-gamma}
\end{eqnarray}
In the momentum space, it can be expressed explicitly as
\begin{eqnarray}
     S_F(p) &=&
\left \{\begin{array}{cc}
       \left ({\rlap\slash p}-\gamma^5 {\cal
             A}_c\chi^c O_{(+)}-\gamma^5 {\cal A}^c\overline\chi_c O_{(-)},
       \right ){\displaystyle i\over
                   \displaystyle p^2-\chi^2} &\hspace{0.8cm}
          \mbox{For quark type I} \\
         & \\
    {\displaystyle i{\rlap\slash p}\over
     \displaystyle p^2} &\hspace{0.8cm}\mbox{For quark type II}
\end{array}
\right .,
\label{Prop-gamma1}
\end{eqnarray}
where a quark of {``type I''} is the one that has color different from
$\chi^c$ (or $\overline\chi_c$) and a quark of {`` type II''} is the
one that has the same color as $\overline\chi_c$ (or conjugate to that
of $\chi^c$). It can be seen that a quark of type II has no gap for
its excitation and an excitation of a quark of type I has a finite gap
$\sqrt {\chi^2}$.

\subsection{Model II and the $\beta$ phase}

The second model Lagrangian density is 
\begin{eqnarray}
{{\cal L}}_2&=& {1\over 2}\overline\Psi \left [ i{\rlap\slash\partial} -
\sigma -
 i\vec{\pi}\cdot\vec{\tau}\gamma^5 O_3 +
\left (\phi^c_\mu\gamma^\mu\gamma^5
{\cal A}_c
  +\vec{\delta}_\mu^c\cdot\vec{\tau}\gamma^\mu{\cal
A}_c \right ) O_{(+)} -  \left (\overline\phi_{\mu
c}\gamma^\mu\gamma^5{\cal A}^c
  +\vec{\overline\delta}_{\mu c}\cdot\vec{\tau}\gamma^\mu{\cal
A}^c \right) O_{(-)}\right ]\Psi \nonumber \\
&&-{1\over 4\widetilde G_0}(\sigma^2 + \vec{\pi}^2) - {1\over
2\widetilde G_{3}}
(\overline\phi_{\mu c}\phi^{\mu c} + \vec{\overline\delta}_{\mu
c}\cdot
\vec{\delta}^{\mu c}),\label{Model-L-2}
\end{eqnarray}
where $\sigma$, $\vec{\pi}$, $\phi^c_\mu$, $\overline\phi_{c\mu}$,
$\vec{\delta}^c_\mu$ and $\vec{\overline\delta}_{c\mu}$ are auxiliary
fields.  It is symmetric under the chiral $SU(2)_L\times SU(2)_R$
group transformation.  This model is discussed in detail in
Refs. \cite{Ying1,Ying11}.  It is written down here for references in
the later sections.  The Euclidean effective potential for this model
used for a determination of the vacuum phase structure in a
Hartree--Fock approximation is found to be
\begin{eqnarray}
V^{eff} &=&
 -{1\over 2}\int^\Lambda
{d^4p\over (2\pi)^4}\left (8\sum_{i=1}^4
ln{\lambda_i(p)
\over\lambda^{(0)}_i(p)} + 4\sum_{i=1}^4
ln{\lambda'_i(p)\over{\lambda'}_i^{(0)}
(p)}\right ) + {1\over 4\widetilde G_0}\sigma^2 + {1\over
2\widetilde
G_3}\overline\phi_\mu\phi^\mu\nonumber\\
&=& -4\int^\Lambda {d^4p\over (2\pi)^4} ln\left [ (1
+{\sigma^2+\phi^2\over
p^2})^2 + {\sigma^2\over p^2}
(1+{\sigma^2\over p^2})(1+{\sigma^2-\phi^2\over p^2})
-4(1+{\sigma^2\over p^2}){(\phi\cdot p)^2\over p^4}\right ]\nonumber\\
&& + {1\over 4 \widetilde G_0}\sigma^2 + {1\over 2\widetilde
G_3}\phi^2,\label{EffV3}
\end{eqnarray}
A numerical evaluation of Eq.~\ref{EffV3} shows that the absolute
minimum of $V^{eff}(\sigma^2,\phi^2)$ is located at either
$\sigma^2\ne 0$ and $\phi^2 = 0$ or $\sigma^2 = 0$ and $\phi^2 \ne 0$
in the spontaneous symmetry breaking phases.  The phase diagram for
this model is presented in Fig. \ref{Fig:bound2}.  Three kinds of
phases for the vacuum are possible. The first phase, which is
identical to the $O$ phase discussed above, is the bare vacuum.  The
second phase, which is the same as the $\alpha$ phase introduced
above, has non-vanishing vacuum expectation value of
$\overline\Psi\Psi$; the chiral $SU(2)_L\times SU(2)_R$ symmetry is
spontaneously broken down to a $SU(2)_V$ flavor symmetry in this
phase. The third phase is labeled as the {\em $\beta$ phase}. It has
non-vanishing diquark and antidiquark condensation characterized by a
non-vanishing $\phi^2$; the chiral symmetry is spontaneously broken
down to a flavor symmetry, in the same way as in the $\alpha$ phase.

The phase transition is of second order across $O$ phase and the
$\alpha$ phase boundary ($\alpha_0=\pi/12$ and $\alpha_3 < \pi/4$).
It is first order phase transition across the $\alpha$ and $\beta$
phase boundary ($\alpha_0>\pi/12$ and $\alpha_3>\pi/4$).  There is a
second order phase transition across the $O$ and $\beta$ phase
boundary ($\alpha_0<\pi/12$ and $\alpha_3 = \pi/4$).

In the $\beta$ phase, the propagators for the quarks are found
\cite{Ying1} to be
\begin{eqnarray}
S_F(p) &=& \left \{ \begin{array}{cc}
           \left ( 1 - iO_2 {\displaystyle {\rlap\slash p}\over
           \displaystyle p^2}{{\rlap\slash\phi}}^c\gamma^5{\cal A}_c
           \right ) i {\displaystyle (p^2-{\phi}^2){\rlap\slash p} - 2
                 p\cdot{\phi}{{\rlap\slash\phi}}
              \over \displaystyle (p^2-\phi^2)^2 -
               4(p\cdot\phi)^2}   \hspace{0.8cm}
               &\mbox{For quark of type I}\\  &\\
          i{\displaystyle {\rlap\slash p}\over
          \displaystyle p^2} \hspace{0.8cm} &
          \mbox{For quark of type II}
\end{array}
\right ., \label{SFp2}
\end{eqnarray}
where $\phi^2 =\overline\phi_{\mu c}\phi^{\mu c}= - \phi^c_\mu\phi^{\mu
  c}$. In order to simplify the computation, a special choice for
the complex phase of the non-vanishing auxiliary fields $\phi_\mu^c$
and $\overline\phi_{c \mu}$ is made, namely, $\overline\phi_{c\mu}$ is chosen to
be $\overline\phi_{c \mu}= -\phi_\mu^c$.

\section{The Vacuum Structure of the $\alpha$ , $\beta$  and
         $\omega$ Phases}
\label{sec:Vacua}

The possible phases of two model Lagrangian densities are studied in
the previous section using the conventional approach to effective
potential in the Hartree--Fock approximation.  A more detailed
characterization of these vacua and their properties is given in the
following using the general framework developed in section
\ref{sec:FDth}.

\subsection{The $\alpha$ phase}

     The effective potential for the $\alpha$ phase in the full theory
given in section \ref{sec:FDth} can be obtained by computing the right
hand side (r.h.s.) of Eq. \ref{Veff-a-FD} using the $p^0$ integration
contour ${\cal C}$ in Fig. \ref{Fig:ConFDB0}, namely
\begin{eqnarray}
    V_{eff}(\mu,\epsilon) &=& 6i\int_{\cal C} {d^4p\over (2\pi)^4} \ln\left (
            1- {\sigma^2\over p_+^2} \right ) \left (
            1- {\sigma^2\over p_-^2} \right )
               + {1\over 4 G_0}\sigma^2 +  {3\over 2\pi^2} \mu^4.
\label{Veff-a-FDF}
\end{eqnarray}

The effective potential has only a trivial minimum located at
$\mu=\epsilon=0$ when the integration contour is chosen to be the
quasiparticle one shown in Fig. \ref{Fig:ConFDB1}. This contour, which
selects only the quasiparticle contributions, violate the conservation
of baryon number explicitly; it is demonstrated in section
\ref{sec:Instab}.  It is necessary to turn the contour ${\cal C}$ into
the complex plane to find the Euclidean stationary points of the field
configurations in a way that preserves the causal structure of
Fig. \ref{Fig:ConFDB0}. Such a contour is also displayed in
Fig. \ref{Fig:ConFDB1}.

A numerical study shows that the minima of $V_{eff}$ for a given value
of non-zero $\sigma$ is located either at $\mu\ne 0$ and $\epsilon=0$
or at $\mu=0$ and $\epsilon\ne 0$. Fig. \ref{Fig:NJ-sec} shows
$V_{eff}$ along three different directions in the $\mu-\epsilon$
plane. The absolute minimum of $V_{eff}$ is
located at $\epsilon=\epsilon_{0}\ne 0$ and $\mu=0$. This is the
result I regard as natural since it avoids the not observed CP
violation associated with the $\alpha$ phase found in section
\ref{sec:Models}. It also agrees with the physical picture for the
$\alpha$ phase, which is condensed with correlating quark--antiquark
pairs.

With a non-vanishing $\epsilon$, Let us first assess the nature of the
primary statistical gauge field excitation.
The effective action for the primary 
statistical gauge field ${\mu'}^\alpha$ at
long distances is given by (see Eq. \ref{Seff-mu2})
\begin{eqnarray}
      S^{(\mu)}_{eff} = \int d^4 x \left [ -{Z^{(\mu)}\over 4} f_{\mu\nu}
      f^{\mu\nu} +  \left ({6\over \pi^2} \epsilon^2 \right )
       {\mu'}^2 \right ].
\label{Seff-mu4}
\end{eqnarray}
Since $\epsilon^2\ne 0$ in the $\alpha$ phase, the $\mu'_\alpha$
excitation is massive (or short ranged) in the static limit and is now
stable against quantum fluctuations.  This agrees also with our
observations since no corresponding long range force and large CP
violation are observed at the present-time condition.

Due to the presence of a finite $\epsilon$, the response of the system
to external (or internal) excitations is different from the familiar
one we have learned from the conventional approaches. When Dirac's
view for the bare vacuum of fermions is taken, namely, the bare vacuum
corresponds to a state in which the negative energy states are filled
and the positive states empty, the vacuum where $\epsilon\ne 0$ is the
state in which all single particle states with energy $E\le-\epsilon$
and $0\le E<\epsilon$ are filled whereas other single particle states
empty. Since for an uniform system, the value of the mass $\sigma$ for
the quasiparticle is always larger than $\epsilon$ in the models
considered, the presence of a finite $\epsilon$ in the vacuum of the
$\alpha$ phase appears to has no effects if the quasiparticle can
propagate long enough without suffering from further scattering. The
presence of $\epsilon$ only provides a virtual possibility for an
uniform system that the local fluctuations of the fields can feel.
So, it has genuine physical effects on local observables even in
uniform systems according to the discussion in Appendix
\ref{app:thedark}.

For a non-uniform system in which the energy of the operator in
Eq. \ref{EigenEq2} could be smaller than $\epsilon$, the presence of
$\epsilon$ may has a real effect on the dynamical processes. For
example, in a chiral soliton in which the energy of the lowest orbits
for the valence fermions moves with the size of the soliton, the
presence of $\epsilon$ will limit the range of change of the soliton's
possible size which gives an extra stability of such solitons.

To put the above argumentation in a more concrete context, let us
consider a situation in which the lowest energy valence fermion state
lies within the region $-\epsilon \le E < 0$ for one size and shape,
then a change in its size or shape that moves $E$ upward can be
continued freely only until $E=0$ since the $0\le E <\epsilon$ states
are filled and the next available state is the one with energy
$E=\epsilon$, which can only be reached by a discontinuous change in
the size and shape of the soliton. If the nucleon can be regarded as a
chiral soliton, this mechanism can prevent it from dissolving inside a
nucleus if the lowest energy valence (constituent) quark states inside
the nucleon lies between ($-\epsilon$,0). Other implications of a
nonvanishing $\epsilon$ are worth to be studied in future works.

Perhaps other interesting implications of a non-vanishing $\epsilon$
are on the particle production and dissipation processes in
non-equilibrium situations like the heavy ion collision.

\subsection{The $\beta$  and $\omega$ phases}

  According to the physical picture discussed so far, the baryon number
content of the $\beta$  and $\omega$ phases ought to be different from
the $\alpha$ phase due to the fact that, instead of quark
and antiquark pairs, quark pairs  and antiquark pairs (or diquark and
antidiquark) are condensed in the vacuum. This is reflected in the
fact that in these two vacua, the expectation value of $\chi^c$
(together with its conjugate field $\overline\chi_c$) and
$\phi^c_\mu$ (together with its conjugate field $\overline\phi_{c\mu}$) are
non-vanishing. 

In the $\beta$ and $\omega$ phases, where $\sigma=0$,
the effective potentials for model
I and II are found, using Eq. \ref{Veff-FDB}, to be of the following
forms
\begin{eqnarray}
  V_{eff} &=& -{4\over (2\pi)^4}\int_{\cal C} d^4p  \ln
              \left (1 + {\chi^4-2\chi^2 p_+\cdot p_-\over p^2_+ p^2_-}
              \right ) + {1\over 2 G_{3'}} \chi^2 \nonumber \\
          & & \hspace{4.5cm} + {3\over 2\pi^2} \left (\mu^4 +
              2\epsilon^4
             + 12 \mu^2\epsilon^2 \right ) \hspace{2cm} \mbox{For Model
               I}\label{Veff-M1}\\
  V_{eff} &=& -{4\over (2\pi)^4}\int_{\cal C}d^4p \ln
              \left (1 + {\phi^4-2\phi^2 p_+\cdot p_-\over p^2_+ p^2_-}
              - 4{(p\cdot\phi)^2\over p_+^2 p_-^2}
              \right ) + {1\over 2 G_{3}} \phi^2 \nonumber \\
          & & \hspace{4.5cm} + {3\over 2\pi^2} \left (\mu^4 +
              2\epsilon^4
             + 12 \mu^2\epsilon^2 \right ) \hspace{2cm} \mbox{For Model
               II}\label{Veff-M2}
\end{eqnarray}
Numerical evaluations show that the local minima of the above
effective potentials are located either at $\mu\ne 0$ and $\epsilon=0$
or $\mu=0$ and $\epsilon\ne 0$. The corresponding $V_{eff}$ in the
$\mu$--$\epsilon$ plane along three different directions are plotted
in Figs. \ref{Fig:SPS-sec} and \ref{Fig:SPV-sec} respectively.  The
absolute minima of both effective potential are located at $\mu \ne 0$
and $\epsilon=0$. That is, in the $\beta$ and $\omega$ phases where
quark pair or diquark condense, the vacua of the systems are the ones
with finite density of baryons with the baryon density given by
Eq. \ref{bar-rho-mu}. 

Such a phase also spontaneously violates the CP invariance of the
system's original Lagrangian density due to the existence of a
non-vanishing CP odd order parameter $\mu^\alpha$ in these phases. In
addition, a pattern in which baryonic matter and antimatter are
separated in space for the $\beta$ and $\omega$ phases of the vacuum
are energetically favored. The superselection sector in the Hilbert
space in which the physical states stays in for such a CP violating
phase can then be determined. Assuming that the statistical ``electric''
field $\pi_u$ is finite, then
Eq. \ref{Q-eigenstate} tells us that $\varsigma=\overline\rho$ when
the spacetime approaches infinity. Due to the translational invariance
of the vacuum state, it is natural to require that
$\varsigma=\overline\rho$ at all spacetime position. In this way the
physical states labeled by $\varsigma$ in the CP violating phase of
the system are determined.
 
    One of the interesting properties of the $\beta$ and $\omega$
phases is that these phases are expected to have off diagonal long
range order ODLRO \cite{ODLO} due to the spontaneous breaking down of
the $U(1)$ symmetry corresponding to baryon number
conservation. Macroscopic quantum phenomena manifest in those phases
possessing ODLRO\footnote{ODLRO is absent in the normal phases of
matter where classical picture is supposed to emerge for macroscopic
systems due to decoherence characterized by a diagonalization of the
effective density matrix of the system interested during the time
evolution (see, for example, Refs. \cite{Decoh}).}. The quantum nature
of these phases leads to behaviors of the system not expected from our
daily experiences. Some of these behaviors are observed in the
superfluid state of $^4$He. Could it allows a quantum mechanical jump
(transition) from an initially zero baryon density $\beta$ or $\omega$
phases of the vacuum state to the lowest energy state of the system
inside relatively large regions: some of them contain baryonic matter
and others contain antibaryonic matter? Such a kind of transition is
forbidden in classical picture since it violates the the relativistic
causality. It is allowed in the quantum measurement processes
according to the standard interpretation of quantum mechanics, in
which a collapse of the wave function of the system occurs after a
measurement.  It remains to be understood in the future more
detailed researches. If it is permitted, then each of these regions
can has a finite size at the moment of the transition or, in another
word, each of them can has a size large than its event horizon.

This property of the $\beta$ and $\omega$ phases could provide a
mechanism for the baryogenesis in the early universe in a
matter--antimatter symmetric universe \cite{Omnes}. It was shown to be
impossible for such a baryogenesis mechanism to be compatible with
observations if the process of baryon--antibaryon separation is
classical \cite{Steigman}. The discovery of the possible $\beta$ and
$\omega$ phases for the strong interaction vacuum in this study might
provide a theoretical basis for a reconsideration of the idea of
matter antimatter symmetric universe. Some of the other more detailed
consequences of this picture, which is beyond scope of this paper, is
worthy of exploring.

The ${\mu'}_0$ degrees of freedom is non-propagating in the $\beta$
and $\omega$ phases due to a non-vanishing (uniform) baryon number
density is present in these two phases.  According to Eq.
\ref{Seff-mu3}
\begin{eqnarray}
      S^{(\mu)}_{eff} &=& {i\over 2}\int d^4 x \left [
      (\overline\rho^2) {\mu'}_0^2 + \ldots\right ].
\label{Seff-mu5}
\end{eqnarray}
It shows that the ${\mu'}_0$ fluctuation is damped by a
non-oscillating Gaussian factor with the width proportional to the
inverse square of the baryon number density in these phases (i.e.,
$\overline\rho^2$). The spatial component of $\mu^\alpha$ in the
$\beta$ and $\omega$ phases is however long ranged. This can also be
realized through an inspection of Eq.  \ref{Seff-mu3}.

The question of whether or not there is a condensation of statistical
mono-poles in these two phases, which is discussed in a general term
in subsection \ref{subsec:more} should be studied in future more
detailed works.

\subsection{The excitations of the primary statistical gauge field}
\label{subsec:abslro}
   In the above discussion it can be seen that in the $\alpha$ ,
$\beta$ and $\omega$ phases, either $\epsilon$ or $\mu$ (it is
equivalent to $\bra{0} j \ket{0}$) is nonvanishing. From the
discussion presented in the subsection \ref{subsec:more}, it can be
concluded that there is no long range order for the time component of
the primary statistical gauge field $\mu^\alpha$ at long distances or
$\mu^0$ corresponds to at most a massive excitation in the non-trivial
vacuum discussed in this paper.

  The situation for the spatial component of $\mu^\alpha$ is
different between the $\alpha$ and $\beta$ or $\omega$ phases. In the
$\alpha$ phase, the excitation related to $\mu^\alpha$ is short
ranged. In the $\beta$ and $\omega$ phases, the excitations related to
\mbox{\boldmath{$\mu$}} are long ranged. These excitations can
therefore generate a statistical ``magnetic force'' between different
particles within the $\beta$ or $\omega$ phases of the vacuum. The
consequences of such a statistical ``magnetic force'' on the evolution
of the system is worthy of studying.

\section{Discussion and Outlook}
\label{sec:Summ}

It is found that at least three interwinding new theoretical elements
are necessary to be brought into a consistent treatment of the
problem. The first one is the general existence of the so called {\em
dark component} in an interacting system originated from the transient
and short distance quantum fluctuations of the system, which is
measured by the difference between the absolute value and the apparent
value of some conserved quantities like the baryon number density,
energy density, etc. of the system. The second one is related to the
recognition of the existence of the so called {\em blocking effects}
in the non-trivial phases of a system. The third one is related to the
necessity of introducing a {\em primary statistical gauge field}
coupled to the fermion (baryon) number current density of the
system. By introducing these three elements into the formulation of
the problem, the door to go beyond the physical pictures limited by
the approximated concept of quasiparticles is open, which allows us to
explore new physical possibilities.

 A systematic path integration formalism for the investigation of the
quantum aspects of an interacting fermion system (or sector) sampled
by Euclidean spacetime stationary configurations in which a
condensation of fermion pairs (fermion pairs, antifermion pairs, and
fermion--antifermion pairs) is present is developed based upon the
asymptotic grand canonic ensemble.  Two statistical parameters, namely
the primary statistical gauge field $\mu^\alpha$ and the statistical
blocking parameter $\epsilon$ are introduced to allow a finer
characterization of the vacuum structure of the system.  In addition,
it is shown that the asymptotic grand canonic ensemble reduces to the
grand canonic ensemble as the spacetime resolution of observation is
sufficiently lowered. Such a behavior is a necessary condition for the
usefulness of describing the macroscopic properties of an interacting
system in terms of particles in certain domain of energy and for the
smooth approach to the well established results in non-relativistic
condensed matter systems at low energies. Combined with the Euclidean
approach to the effective action, some of the quantum effects that
survive the thermodynamic limit can be included. The present approach,
which uses quantum field theoretical language, is consistent with
thermodynamics and can be extended to finite temperature case
\cite{TFTQFT}.

Firstly, the dark component for local observables like the fermion
number density does exist in interacting theories in which the direct
association of the field theoretic definition of fermion number
density with the number of ``free particles'' per unit volume becomes
obscure especially when a phase transition inside the system has
occurred.  This conclusion is also applicable to other local
observables like the energy density, which may have
implications on the dark matter problem in Cosmology since it implies
that the apparent matterless space at the macroscopic level is capable
of revealing itself of matter effects in low energy gravitational
processes when $\mu$ is below the baryonic particle production
threshold even after the energy density for the $\mu=0$ state is
subtracted. This is because gravitational fields couple {\em locally}
to the source matter fields which contain the random quantum
fluctuation generated dark component. Further researches in that
direction in the context of understanding the cosmological baryogenesis and 
dark matter problem is an interesting direction to be explored.

In addition, the picture that a nucleon is made of three valence
quarks (quasiparticles) need to be modified when there is additional
close-by virtual phases for the hadronic vacuum state that has
slightly higher energy density than the actual one. The implications
of such a finding can be explored in observables related to a
nucleon. Some of them are studied in Refs. \cite{PCACsuc,GDHp}, other
related problems concerning a nucleon, like the understanding of the
origin of the Gottfried sum rule violation in deep inelastic
scattering confirmed in a recent measurement \cite{GFsum}, the small-x
behavior of nucleon structure functions in deep inelastic scattering
\cite{F2-pap,g1-pap}, etc..

Secondly, it is important to take into account the fermionic blocking
effects due to the presence of a macroscopic population of bare
particles in the nontrivial vacuum phases of a system. The blocking
effects are generally included in the theory by introducing the
statistical blacking parameter $\epsilon$, which is non-zero for a
system's certain vacuum phases.

The effects of the blocking can not be progressively generated using
perturbative expansion starting from a field configuration with 
$\epsilon=0$. In the $\alpha$ phase of the strong interaction vacuum
discussed here, $\epsilon=0$ configurations are inconsistent ones
since it is known that the $\alpha$ phase of the strong interaction
vacuum is macroscopically populated with the current quarks and
antiquarks, which changes the available states for an current quark due
to Pauli principle. The discovery of the blocking effects has hitherto
unnoticed implications related to, e.g., the stability of a nucleon in
a nucleus and nuclear matter, the mechanism for particle production in
a heavy ion collision, new ways of (quasi)particle dissipation in a
strong interacting system, etc.. 

Thirdly, the statistical gauge degrees of freedom of the system
represents certain collective mode of the system that has a dynamics
of its own. There are two components for the primary statistical gauge
field: the first one is the classical configurations which serves as a
background field; the second one is the local fluctuations of it
around the classical configurations. 

The classical configurations, which is a spacetime independent
background $\mu^\alpha$ in the case studied here, play the role of the
chemical potential in the the conventional non-relativistic
approach. It determines the asymptotic grand canonic ensemble of the
system. It can also has non-trivial topological configurations
corresponding to different quantized statistical ``magnetic flux''
which can determine the phase structure of the system on a finer
basis. Once present, the statistical ``magnetic field'' affects the
dynamical evolution of the system that can result in 1) material
pattern formation and 2) providing the seed for the galactic magnetic
field in the early universe. Whether or not such an idea is actually
relevant to comprehend what happened in the the early universe can be
studied in further works.

The local fluctuations of it around the background extended
configuration represent the corresponding dynamical excitations of the
system.  The necessary condition for the existence of long range
statistical gauge correlation in various possible phases of the system
is discussed.

The statistical gauge degrees of freedom are also introduced in
condensed matter physics in the context of the half filled Hubbard
model, which serves as one of the prefered models that is expected to
describe the phenomena of ``high temperature superconductivity'' in
certain matterials \cite{Affleck,Kotliar}. The motiviation for
introducing the statistical gauge degrees of freedom there is quite
different from the ones in this work. Here, the statistical gauge
degrees of freedom are introduced {\em a priori} based upon locality
and Lorentz invariance with the intention of describing the
relativistic fermionic systems. Since the approach in this paper is
applicable to all fermionic system, it is quite interesting to see
whether the non-relativistic reduction of the problem can lead to some
form of statistical gauge degrees of freedom for condensed matter
systems at low energies, including the ones that Hubbard model
describes. Nevertheless, many techniques in treating the statistical
gauge degrees of freedom in condensed matter physics are expected to
be applicable or a least adaptable here.

One of the differences at formal level between the approach here and
the ones used in condensed matter physics manifests in the different
criterion for the selection of physical states within the full
representing Hilbert space of the problem.  The physical states in the
statistical gauge theories developed so far in condensed matter
physics is invariant under infinitesimal local gauge transformations,
which is realized by the requirement that the operator form of the
``Gauss law'' annihilates all physical states in the superselection
sector of the Hilbert space \cite{Fradkin}. Such a strict enforcement
of the statistical gauge invariance on the physical state vector of
the system is neither necessary nor desirable for the statistical
gauge invariant systems since it would exclude all finite density
state from the physical sector of the system if one requires that the
statistical ``electric field'' (denoted by $\pi_u$ here) is finite. Such
a situations is certainly unacceptable. Albeit this embarrassment can
be circumvented in the 2+1 dimension \cite{Fradkin}, it is not
expected to be easily implemented in higher dimensions. For the
statistical gauge transformations, the statistical gauge invariance
can be implemented by a imposing a less restrictive conditions on the
physical superselection sector of the Hilbert space. Instead of
requiring that the physical states are invariant under the gauge
transformation, the gauge invariance on observables can be implemented
by requiring that all states in a physical superselection sector of
the Hilbert space change a (coordinate dependent) {\em common
phase}. The mathematical form for such a requirement is represented by
Eq. \ref{Q-eigenstate}. The physical superselection sectors are then a
functional determined by the common function $\varsigma$. With such a
generalization of the ``Gauss law'' for the theory, both the
requirements of the finiteness of the statistical ``electric field''
$\pi_u$ and of the fact that finite density states are actually
physical states can be met consistently.

The formalism is then applied to two half bosonized model Lagrangian
densities. Four possible phases for the vacuum state of the
interacting relativistic chiral symmetric systems are found. The first
phase, called the $O$ phase, correspond to the bare vacuum state of
the system. Fermion--antifermion pairs condense in the second phase, 
named the $\alpha$ phase, of its vacuum. The $\alpha$ phase has the
following properties: 1) the chiral $SU(2)_L\times SU(2)_R$ symmetry
of the Lagrangian density of the system is spontaneously broken down
to a $SU(2)_V$ symmetry 2) the baryon number density is zero 3)
statistical  blocking effects exists 4) the statistical gauge
correlation is short ranged due to the presence of the 
statistical blocking effects. The third and fourth possible 
phases of the vacuum are called the $\omega$ phase and $\beta$ phase
respectively. Fermion pairs and antifermion pairs condense in the 
$\omega$ and $\beta$ phases. It is found that in these two phases of
the vacuum: 1) the original chiral $SU(2)_L\times SU(2)_R$ symmetry of
the Lagrangian density of the system remains unbroken in the $\omega$
phase and is spontaneously broken down to a $SU(2)_V$ symmetry in the
$\beta$ phase 2) the baryon number density is different from zero or
can be locally generated by separating fermion and antifermion rich
region spontaneously 3) the $U(1)$ symmetry corresponding to
electromagnetism is spontaneously  broken down to generate ``massive
photon'' excitations (see \cite{Ying11}) 4) no statistical blocking
effects in these two phases 5) the spatial components of the statistical
gauge excitation is long ranged; the quantum fluctuation in the time
component of the primary statistical gauge field is Gaussian damped 6) off
diagonal long range order exists in these two phases to give rise to
macroscopic quantum behavior for the system, which is suppressed in the
normal phase of the system.

The implication of the finding presented in this paper on physical
processes of strong interaction phenomena that are  currently being or
going to be observed or are in need to be explained theoretically
remains to be investigated in the future. 

\vspace{0.8cm}
\noindent
{\bf Acknowledgment} This work is supported in part by grants from 
        the Post Doctoral Fund of China, 
        the Young Researcher Fund of Fudan University and 
        a Research Fund from the State Education Commission of China.

\newpage

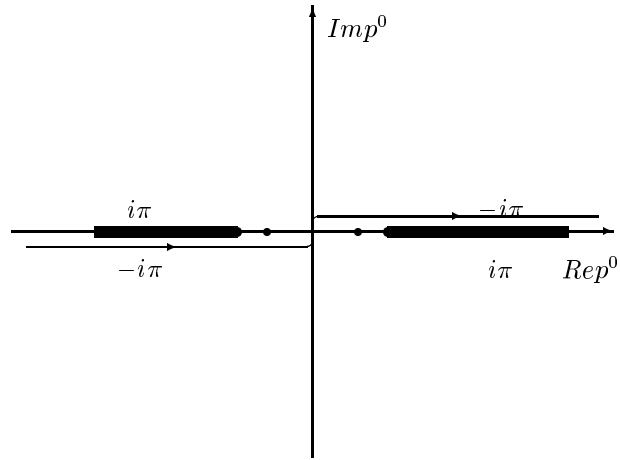
\begin{figure}[h]
\caption{\label{Fig:ConCon1} The $p^0$ integration contour for the
  effective action. Here the filled circle represents the possible
  discrete spectra of $S_F^{-1}[f]$ and the thick lines extending to
  positive and negative infinity represent the branch cuts of the
  logarithmic function. The $\pm i\pi$ above or under the thick lines
  denote the imaginary parts of the logarithmic function on the
  physical sheet of the $p^0$ plane.}
\end{figure}
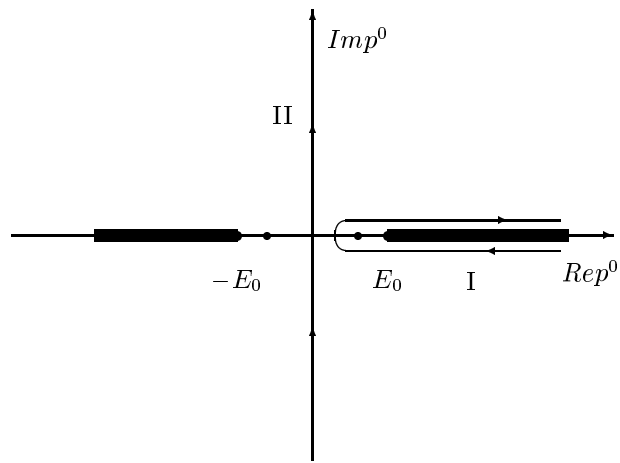
\begin{figure}[h]
\caption{\label{Fig:ConCl2} Contour I is commonly used in literature
  where a non-covariant cutoff in 3-momentum space is applied. It is
  called the quasiparticle contour in this paper. Contour II is the
  one for the Euclidean effective action in the conventional
  approach.}
\end{figure}
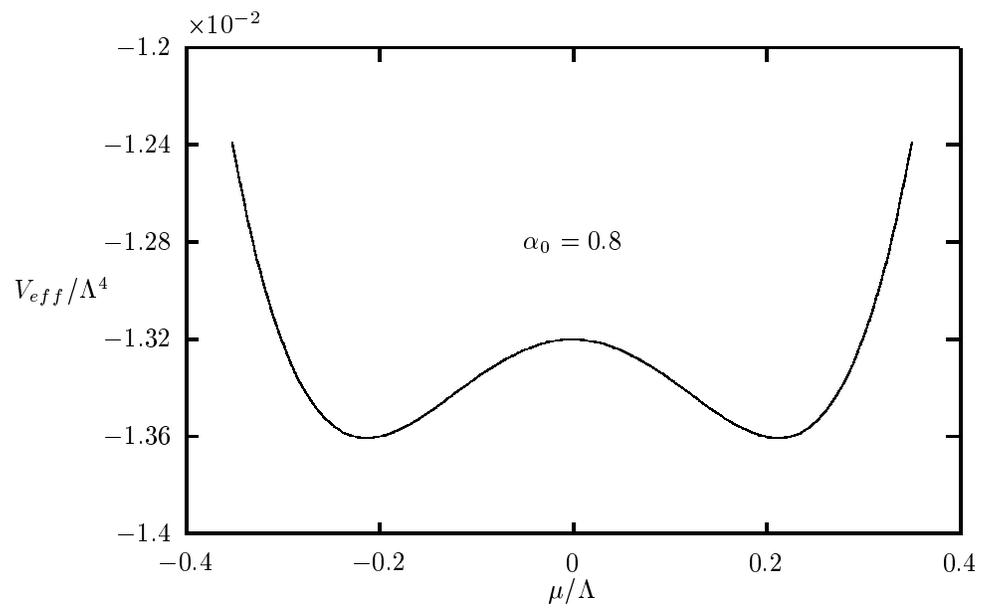
\begin{figure}[h]
\caption{\label{Fig:a-phase-mu} The $\mu$ dependence of $V_{eff}$ 
  of the conventional approach in the $\alpha$ phase. It has two
  minima at non-zero $\mu$. Here $\Lambda$ is the Euclidean momentum
  cutoff that defines the model.}
\end{figure}
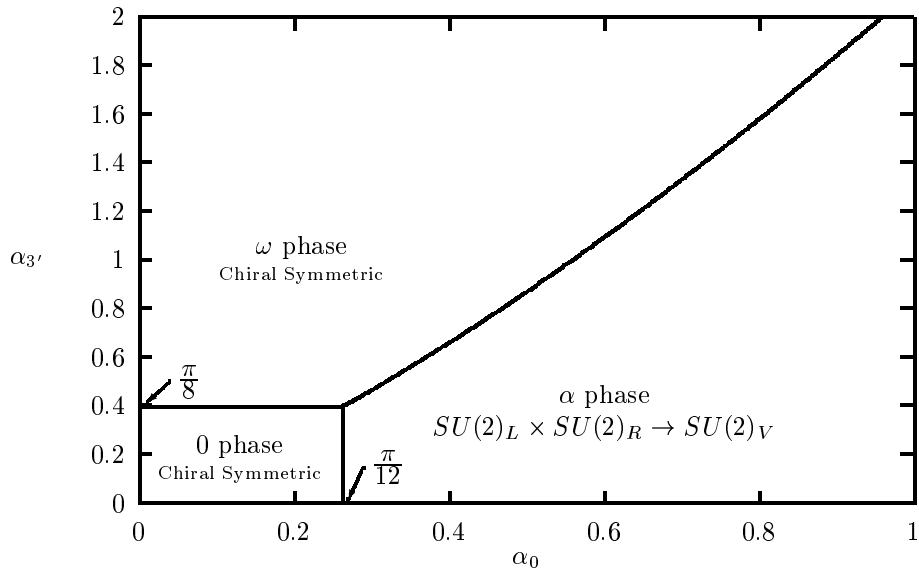
\begin{figure}[h]
\caption{\label{Fig:bound1} The phase boundaries between the
  $\alpha$, $\omega$ and the $O$ phases. The chiral symmetry is
  unbroken in both the $\omega$ phase and the $O$ phase. The $\alpha$
  phase breaks the chiral symmetry spontaneously down to a flavor
  symmetry.}
\end{figure}
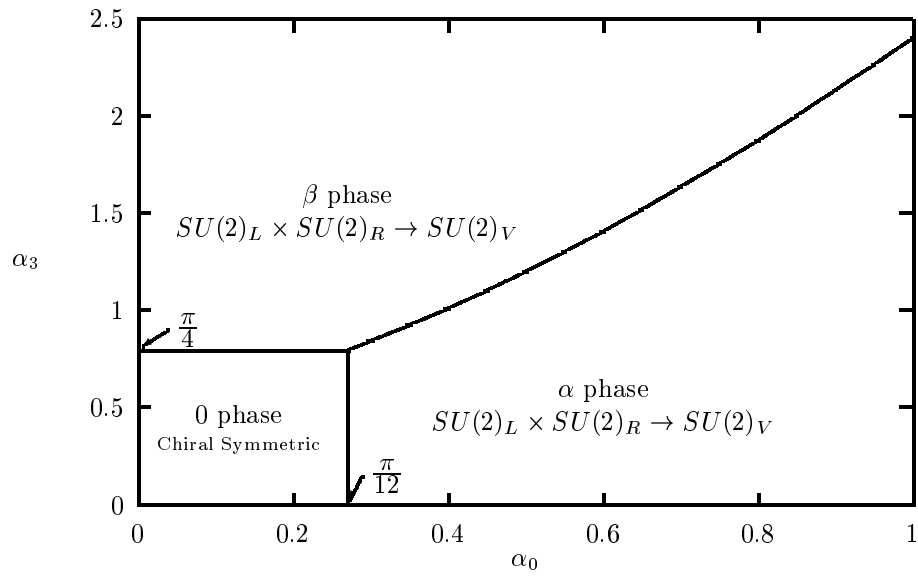
\begin{figure}[h]
\caption{\label{Fig:bound2} The phase boundaries between the
  $\alpha$, $\beta$ and $O$ phases. The chiral symmetry is unbroken in
  the $O$ phase. The $\alpha$ phase and $\beta$ phase break the chiral
  symmetry spontaneously down to a flavor symmetry.}
\end{figure}
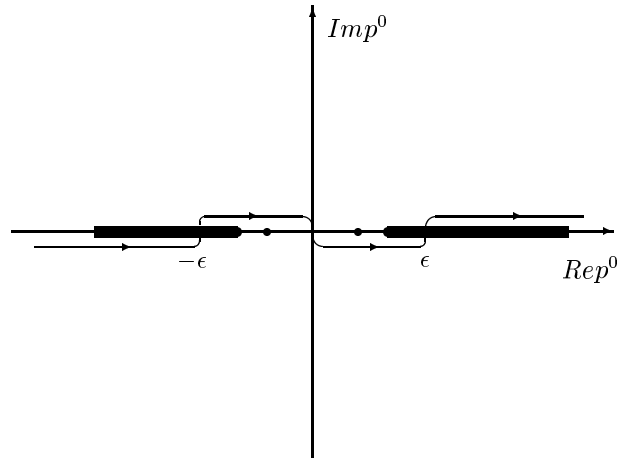
\begin{figure}[h]
\caption{\label{Fig:ConFDB0} The $p^0$ integration contour for the
  effective action for transition amplitudes between states in which
  both the fermion and antifermion states with absolute value of their
  energy below $\epsilon$ filled. Here the filled circle represents
  the possible discrete spectra of $S_F^{-1}[f]$ and the thick lines
  extending to positive and negative infinity represent the branch
  cuts of the logarithmic function. It also represents the $p^0$
  integration contour of the full theory in its original Minkowski
  spacetime form.}
\end{figure}
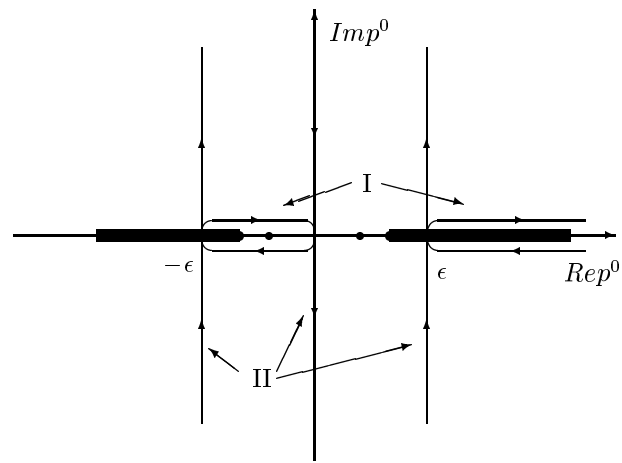
\begin{figure}[h]
\caption{\label{Fig:ConFDB1} The quasiparticle $p^0$ integration contour for
  the full theory is labeled by ``I''. The Euclidean $p^0$ integration
  contour for the full theory, which preserves the causal structure of
  the original one, is labeled by ``II''. }
\end{figure}
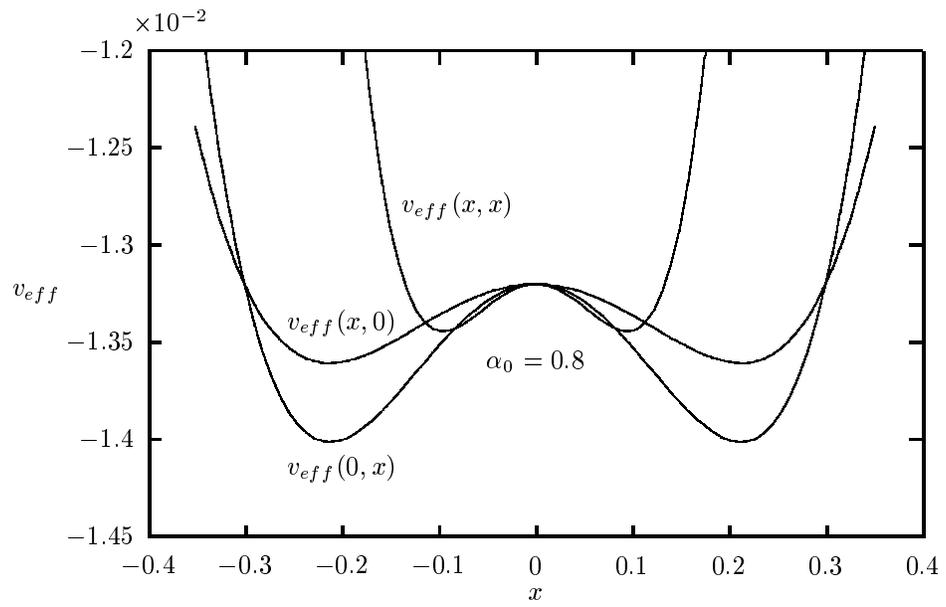
\begin{figure}[h]
\caption{\label{Fig:NJ-sec} The dependences of $V_{eff}(\mu,\epsilon)$
  in the $\alpha$ phase on $\mu$ and $\epsilon$ along different
  directions in the $\mu$--$\epsilon$ plane. The direction in which
  $V_{eff}$ has the smallest value is in the $\mu=0$ direction. Here
  $v_{eff}=V_{eff}/\Lambda^4$ and $x$ variable is either $\mu/\Lambda$
  or $\epsilon/\Lambda$.}
\end{figure}
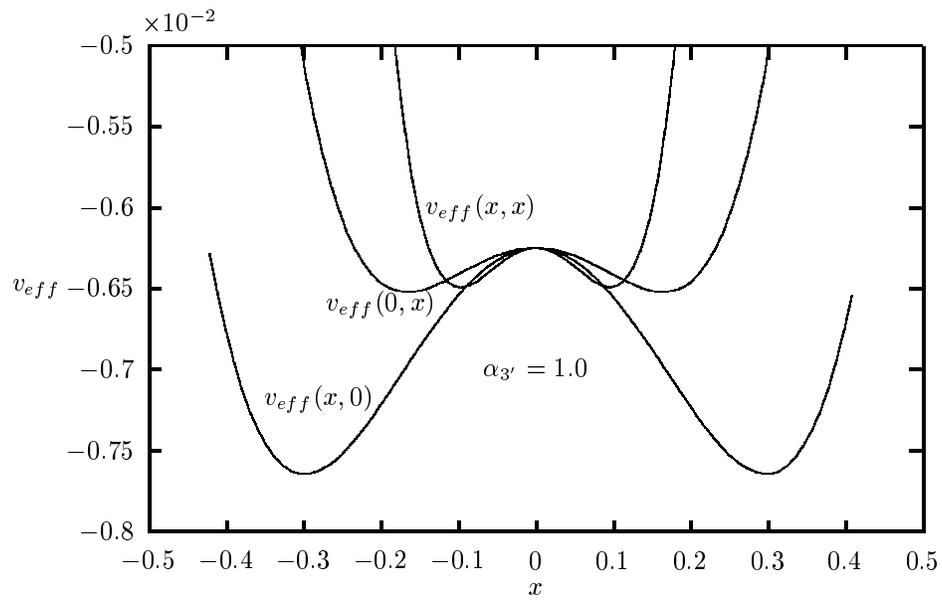
\begin{figure}[h]
\caption{\label{Fig:SPS-sec} The dependences of
  $V_{eff}(\mu,\epsilon)$ in the $\omega$ phase on $\mu$ and
  $\epsilon$ along different directions in the $\mu$--$\epsilon$
  plane. The direction in which $V_{eff}$ has the smallest value is in
  the $\epsilon=0$ direction. Here $v_{eff}=V_{eff}/\Lambda^4$ and $x$
  variable is either $\mu/\Lambda$ or $\epsilon/\Lambda$.}
\end{figure}
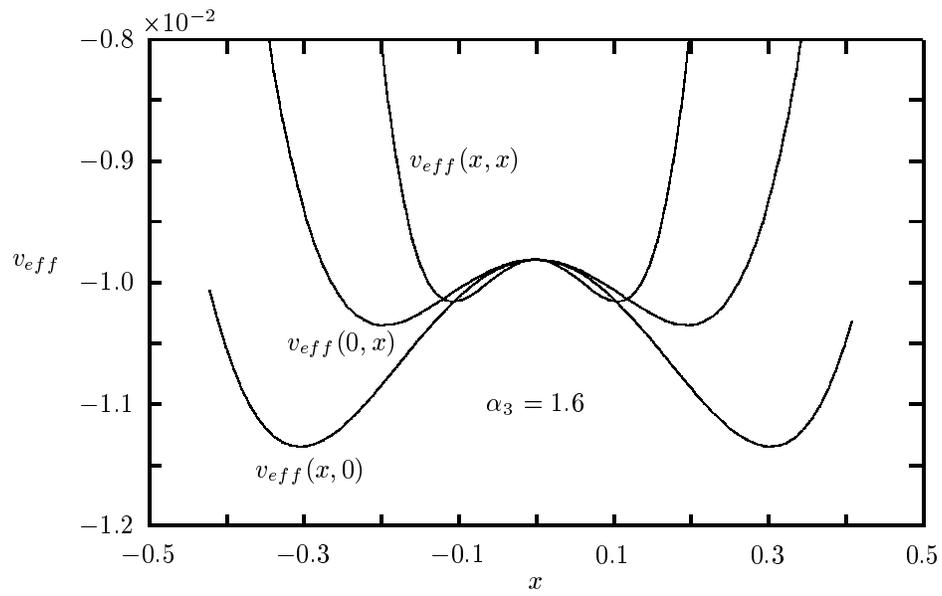
\begin{figure}[h]
\caption{\label{Fig:SPV-sec} The dependences of
  $V_{eff}(\mu,\epsilon)$ in the $\beta$ phase on $\mu$ and $\epsilon$
  along different directions in the $\mu$--$\epsilon$ plane. The
  direction in which $V_{eff}$ has the smallest value is in the
  $\epsilon=0$ direction. Here $v_{eff}=V_{eff}/\Lambda^4$ and $x$
  variable is either $\mu/\Lambda$ or $\epsilon/\Lambda$.}
\end{figure}
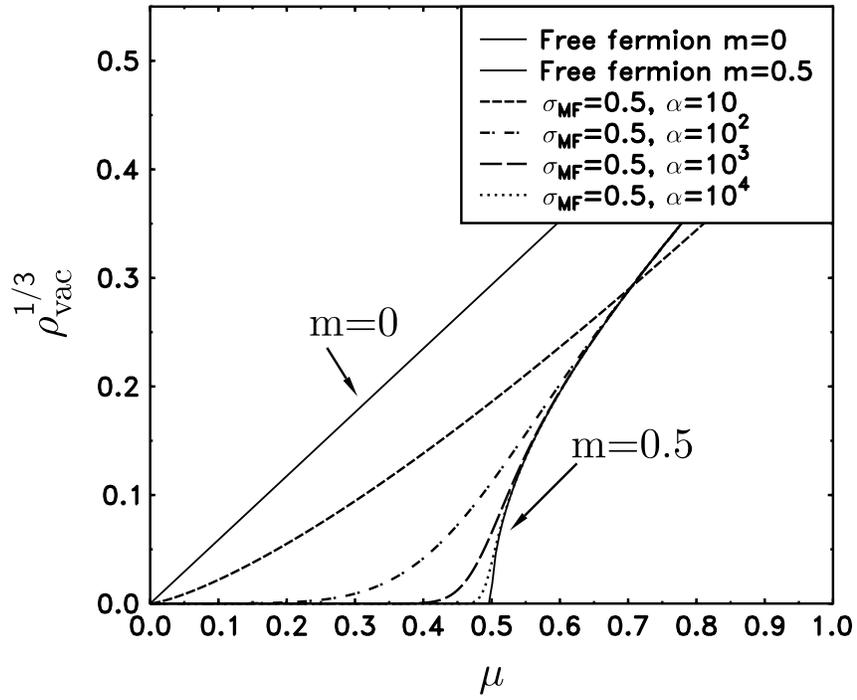
\begin{figure}[h]
\caption{\label{Fig:Fdensity} The dependence of $\rho_{vac}^{1/3}$ on the 
spacetime independent part of the primary statistical gauge field
$\mu$. Here $\alpha =\varpi A$. The unit for the dimensional
quantities are $GeV$. Solid lines represent the case of free theory
with mass 0 and 0.5 respectively. Other lines represent the results
for the massless NJL model with different strength of local
fluctuations characterized by the $\alpha$ of the order parameter.}
\end{figure}

\newpage

\begin{appendix}

\section{Free Fermion Systems with \lowercase{$n_f$} Flavors and
  \lowercase{$n_c$} Colors}
\label{app:FF}

\subsection{With primary statistical gauge field only}

The Lagrangian density for a massive fermionic system with the primary
statistical gauge field $\mu^\alpha$ (see Eq. \ref{General_W_FD})
included has a form
\begin{eqnarray}
   {\cal L} &=& {1\over 2}\overline\Psi \left (i\rlap\slash\partial
                      + \rlap\slash\mu O_3 - m \right )\Psi
,\label{Free_L1}
\end{eqnarray}
where $m$ is the mass of the fermion.

The generating functional $W[\overline\eta,\eta,\mu]$ for such a
system can be written as
\begin{eqnarray}
    e^{W[\overline\eta,\eta,\mu]} &=& \int D[\Psi]
                      e^{i\int d^4x \left ({\cal L} + \overline\eta\Psi +
                        \overline\Psi\eta \right )}= const\times e^{{1\over 2}SpLn\gamma^0 iS_F^{(0)-1}
                        + {1\over 2} \overline\eta S_F^{(0)}\eta}
\label{Free_W}
\end{eqnarray}
with $\overline\eta$ and $\eta$ Grassmann external fields and
\begin{eqnarray}
  iS_F^{(0)-1} &=& i\rlap\slash\partial
                      + \rlap\slash\mu O_3 - m.
\end{eqnarray}
 This equation
allows us to write
\begin{eqnarray}
       iW[0,0,\mu] &=& {i\over 2} SpLn\gamma^0iS_F^{(0)-1} + const,
\label{W00}
\end{eqnarray}
which for an uniform $\mu^\alpha=(\mu,0)$, takes the following
form
\begin{eqnarray}
      iW[0,0,\mu]/V_4 &=&-i n_f n_c\int_{\cal C} {d^4 p\over (2\pi)^4}
                      \ln (p^2_+ - m^2) (p^2_- - m^2) + const,
\label{W00-1}
\end{eqnarray}
where $V_4 \equiv L^3 T$ with $L^3\to\infty$ the spatial
volume and $T\to\infty$ the temporal extension of the system, 
$p_+^\mu = (p^0+\mu,{\bf p})$,
$p_-^\mu = (p^0-\mu,{\bf p})$ 
and ${\cal C}$ denotes the $p^0$ integration contour in the complex $p^0$
plane shown in Fig. \ref{Fig:ConCon1}. 
$W[0,0,\mu]$ can be further specified by requiring
$W[0,0,0]=0$. Such a $W[0,0,\mu]$ is
\begin{eqnarray}
      iW[0,0,\mu]/V_4 &=&-i n_f n_c\int_{\cal C} {d^4 p\over (2\pi)^4}
                      \ln {(p^2_+ - m^2) (p^2_- - m^2)\over (p^2-m^2)^2}.
\label{W00-2}
\end{eqnarray}

Since for free fields, the asymptotic grand canonic ensemble is the
grand canonic ensemble and the quasiparticle approximation is
actually an exact one (Appendix \ref{app:thedark}), the integration
contour for $p^0$ integration above can be chosen as the one shown in
Fig. \ref{Fig:ConCl2} \cite{TFTQFT}. Since the quantum fluctuations of
the free fields are also exactly known and included already, they
should not be sampled using the Euclidean approach.  To avoid over
counting of the quantum fluctuations of the free field, the $p^0$
integration should be done first and then the spatial component of
$p^\mu$. The adoption of the 8-component spinor for the fermion field
makes it equivalent whether the $p^0$ integration is carried out on
the real axis or is on the imaginary axis \cite{TFTQFT} provided that
it is done first.  The result is
\begin{eqnarray}
     iW[0,0,\mu]/V_4 &=& {n_f n_c\over \pi^2}
                  \int_0^{\mu}
                dp p^2 (\mu - E_p) =
                       \mu \overline \rho - \overline \varepsilon 
                  = - \Omega/V_4,
\label{W00-3}
\end{eqnarray}
where $\overline \varepsilon$ is the internal energy density and
$\overline\rho$ is the fermion number density of the system and
$E_p=\sqrt{p^2+m^2}$.  It is evident that $i W[0,0,\mu]$ correspond to
the negative of the grand-potential $\Omega(\mu)$ at zero temperature
in a many body system.

  Before ending of this subsection, it is worth mentioning that had we
adopted the usual 4-component representation for a Dirac spinor, we
would not have obtained Eq. \ref{W00-3} by using the path integration
method with the primary statistical gauge field developed
here. Part of the reasons is due to the non-symmetric way of
introducing the primary statistical gauge field in the 4-component
representation.  This issue is discussed in \cite{TFTQFT} in more
details.

\subsection{The full theory}

   According to section \ref{sec:FDth},
the generating functional $W[\overline\eta,\eta,\mu,\epsilon]$
for an uniform free fermion
system in the full theory can be written as
\begin{eqnarray}
    iW[\overline\eta,\eta,\mu,\epsilon]/V_4 &=&
-i n_f n_c\int_{f.q.p.} {d^4 p\over (2\pi)^4}
                      \ln (p^2_+ - m^2) (p^2_- - m^2) + const
\label{W00-f1}
\end{eqnarray}
with ``$f.q.p.$'' denoting the full
quasiparticle $p^0$ integration contour in the complex $p^0$
plane shown in Fig. \ref{Fig:ConFDB1}. The constant in the above equation
is so chosen that $W[0,0,0,0]=0$, which means that
\begin{eqnarray}
      iW[0,0,\mu,\epsilon]/V_4 &=&-i n_f n_c\left [\int_{f.q.p.} 
                   {d^4 p\over(2\pi)^4}
                      \ln (p^2_+ - m^2) (p^2_- - m^2) -2 \int_{q.p.}
                       {d^4 p\over (2\pi)^4}\ln (p^2-m^2) \right ],
\label{W00-f2}
\end{eqnarray}
where ``$q.p.$'' denoting the quasiparticle $p^0$ contour shown in Fig.
\ref{Fig:ConCl2}. It is found, after some algebra, that
\begin{eqnarray}
      iW[0,0,\mu,\epsilon]/V_4 &=& \left (\mu_+ \overline
            \rho_{(+)} + \mu_- \overline\rho_{(-)}
                             - \mu \overline \rho \right ) - \left (
                          \overline \varepsilon_{(+)} + \overline 
                          \varepsilon_{(-)} -
                          \overline \varepsilon \right ),
\label{W00-f3}
\end{eqnarray}
where $\mu_\pm = \mu \pm \epsilon$ and
\begin{eqnarray}
       \overline\rho_{(\pm)}  &= & {n_f n_c\over \pi^2} \int^{\mu_\pm}_0
       d|\mbox{\boldmath $p$}| |\mbox{\boldmath $p$}|^2,
\label{rho-b-pm}\\
       \overline \varepsilon_{(\pm)} &=& {n_f n_c\over \pi^2} \int^{\mu_\pm}_0 
        d|\mbox{\boldmath $p$}| |\mbox{\boldmath $p$}|^2 E_p.
\label{bar-e-pm}
\end{eqnarray}

\section{The Existence of the Dark Components for Local Observables}
\label{app:thedark}
\subsection{Cluster decomposition and the origin of the dark component}

For simplicity, the two flavor half bosonized NJL model (see Appendix 
\ref{app:NJL}) is used for our discussion. The model Lagrangian density 
with a primary statistical gauge field $\mu^\alpha(x)$ is
\begin{eqnarray}
{\cal L} &=& {1\over 2} 
\overline\Psi\left (i\rlap\slash\partial + \rlap\slash\mu O_3 -\sigma -
i\gamma^5 \vec{\tau}\cdot\vec{\pi} \right )\Psi - {1\over 2 G_0} \left
(\sigma^2 + \vec{\pi}^2 \right ). \label{Model-L-app}
\end{eqnarray}

After performing the path integration over the fermion fields $\Psi$ and
$\overline\Psi$, the generating functional $W[J]$ can be written as
\begin{eqnarray}
    e^{W[J,\mu]} &=& \int D[\sigma,\vec{\pi}]
     e^{iS_{eff}[\sigma,\vec{\pi},\mu] + i \int d^4x f\cdot J},
\label{Gen-Func2}
\end{eqnarray}
where ``$J$'' and ``$f$'' represent, collectively, the external fields and
auxiliary fields respectively. The effective action $ S_{eff}$ is given by
\begin{eqnarray}
S_{eff}[\sigma,\vec{\pi},\mu] &=& - i{1\over 2} Sp Ln
    S_F^{-1}[\sigma,\vec{\pi},\mu] S_F[0,0,0]+{1\over 2G_0} \int d^4 x \left (
            \sigma^2 + \vec{\pi}^2 \right),
\label{Seff-app}
\end{eqnarray}
where $Sp$ denotes the functional trace. The operator
$iS_F^{-1}[\sigma,\vec{\pi},\mu]= i\rlap\slash\partial+\rlap\slash\mu 
  O_3 -\sigma -
i\gamma^5 \vec{\tau}\cdot\vec{\pi}$ is the inversed propagator of the
fermions in the background auxiliary fields $\sigma(x)$ and $\vec{\pi}(x)$.

Let us define $\rho[\sigma,\vec{\pi},\mu;x]\equiv \mbox{Tr}\gamma^0
\mathopen{\langle x\,|}S_F[\sigma,\vec{\pi},\mu] \mathclose{|\,x\rangle} $,
where the trace ``Tr'' is over the internal degrees of freedom of the
fermions. The fermion number density of the model in vacuum state is
\begin{eqnarray}
\sum_{\{f\}}{\cal W}[f,f]
\mathopen{\langle f,t=+\infty\,|}\widehat\rho(x)
\mathclose{|\,f, t=-\infty \rangle}
&=& {1\over i} {\delta \ln Z \over \delta \mu^0(x)}=
           {1\over Z}\int D[\sigma,\vec{\pi}]\rho[\sigma,\vec{\pi},\mu;x]
e^{iS_{eff}[\sigma,\vec{\pi},\mu]},
\label{rho-general}
\end{eqnarray}
where $Z= \int D[\sigma,\vec{\pi}]e^{iS_{eff}[\sigma,\vec{\pi},\mu]}$
and $f$ denotes the collection of $\{\sigma,\vec{\pi}\}$ and ${\cal
W}[f',f]$ is the weight functional of the asymptotic grand canonic
ensemble discussed in the main text. The Minkowski spacetime is not
suitable to study the properties of the vacuum state using
Eq. \ref{rho-general} since the initial and final auxiliary field
configurations are not specified.  The usual procedure to project out
the contributions of the vacuum state is go to the Euclidean spacetime
in which the vacuum state has lowest energy.

In the mean field approximation, the vacuum phase is determined by
minimizing the effective potential $ V_{eff}(\sigma,\mu) = -
S_{eff}[\sigma,0,\mu]/\Omega$ with $\Omega\to\infty $ the spacetime
volume of the system, $\sigma$, $\mu^\alpha$ spacetime independent and
$\vec{\pi}$ assumed zero. Due to the chiral symmetry, the assumption
$\vec{\pi}=0$ does not result in a loss generality. The phase of the
system is determined by the condition $\delta V_{eff}(\sigma,\mu
)/\delta\sigma =0$. The solution for $\overline\sigma$ is non-zero
after the coupling constant $G_0$ is greater than a critical value
$G_{0c}$. A non-vanishing $\overline\sigma$ generates an effective
mass for the fermions, which act as  quasiparticles.

The mean field fermion number density for the vacuum is obtained from
Eq. \ref{rho-general} by ignoring the functional integration over
$\sigma$ and $\vec{\pi}$ and let $\sigma=\overline\sigma$. The result
is (see Appendix \ref{app:FF})
\begin{eqnarray}
\overline\rho_{MF} &=&
\rho(\overline\sigma,0,\mu;x) = {n_c n_f\over 3\pi^2} 
                                      \theta(\mu-\overline\sigma)\left
[\mu^2-\overline\sigma^2\right ]^{3/2},
\label{rhoMF}
\end{eqnarray}
 which is non-zero only when $\mu>\overline\sigma$ just like the
fermion number density of free massive particles with mass
$m=\overline\sigma$.  Such a behavior of $\overline\rho$ is also
predicted in the finite density field theory based on a global
chemical potential $\mu_{ch}$ with $\overline N_{app}$ discussed in
the introduction behave in the same way as $\Omega\overline
\rho_{MF}$. This will be discussed in the following.  Therefore, the
quasiparticle contributions saturate the fermion number density in the
field theoretical approach to finite density problems based on a
global chemical potential.

The contributions of quantum fluctuations around the mean field
$\overline \sigma $ are formally included in
Eq. \ref{rho-general}. The results are commonly expressed as loop
corrections to the fermion number density vertex, which is not
attempted here.

Instead of performing a loop expansion computation of the fermion
number density, the effects of the quantum fluctuations can be
evaluated non-perturbatively by ``doing'' the path integration.

To proceed, the system under consideration is first putted in a
Euclidean spacetime box of length $L$ in each direction and with
periodic boundary conditions at its boundary surfaces. The
thermodynamic limit is defined as the limit of $L\to\infty$.  In the
thermodynamic limit, the extremal configuration dominates the path
integration among those configurations of $\sigma$ and $\vec{\pi}$
that give divergent action in the thermodynamic limit. The
contributing finite action quantum fluctuation configurations, the
number of which is proportional to the spacetime volume $\Omega=L^4$,
are further classified into two categories: 1) correlated localized
configurations, which are defined as the ones that approaches to the
mean field configurations in the spacetime infinity and 2) correlated
extended configurations, which are the ones that remain different from
the mean field configurations at the spacetime infinity.

For a given system, whether or not a configuration is a correlated extended
configurations or is of localized ones\footnote{In the sense that it can be
decomposed into localized ones.} is determined by dynamics. The on-shell
amplitudes, are solutions of the ``classical equation of motion'' in the
Euclidean spacetime
\begin{eqnarray}
{\delta S^E_{eff}[f]\over\delta f(x) } &=& 0,
\label{Eq-onshell}
\end{eqnarray}
with $S^E_{eff}$ the Euclidean effective action, $f$ representing $\sigma$ or
$\vec{\pi}$ fields. The superscript $E$ shall be suppressed in the following.
The set of the extended solutions to Eq. \ref{Eq-onshell} are correlated
ones. Albeit there are plenty of extended on-shell amplitudes in the
Minkowski spacetime, there is no known one in the Euclidean one. I shall
assume the absence of them. The degree of correlation of an arbitrary
extended configuration at different spacetime points is determined by the
degree of their deviation from the extended on-shell amplitudes for
propagating excitations of the system.

The correlation between the off-shell configurations at two different space
time points decreases exponentially. For example, consider the field--field
correlations or propagators
\begin{eqnarray}
    \mathopen{\langle 0\,|}T \widehat f(x_1) \widehat f(x_2) 
    \mathclose{|\,0\rangle} &=&
    \int {d^4 p\over (2\pi)^4} e^{-ip\cdot (x_1-x_2)} {1\over p^2+m^2}
\label{E-fprop}
\end{eqnarray}
with $T$ denoting time ordering. The off-shellness of these configurations is
measured by their mass $m$. The correlation of these configurations at two
different locations ${x}_1$ and $ {x}_2$ (in the Euclidean spacetime)
decreases as fast as $\exp(-r m)/r$ with $r=|x_1-x_2|$. So they can be 
decomposed
into a superposition of localized ones with sizes of order $1/m$. For these
set of configurations, one can divide the spacetime into cells with a
dimension sufficiently larger than their correlation length. Then the
contribution of this set of field configurations to the partition functional
Eq. \ref{Gen-Func2} can be cluster decomposed to
\begin{eqnarray}  Z[J] &\approx & \prod_k
 z_k[J],\hspace{0.4cm} W[J] \approx \sum_k  w_k[J]
     \label{Cluster-Decomp}
\end{eqnarray}
with $ w_k[J]\equiv \ln  z_k[J]$, $ z_k[J]$ and $ w_k[J]$ the corresponding
partition functional of the kth cell. The full partition of the kth cell is
\begin{eqnarray}  z_k[0] &\sim & \int D_k[f] e^{-S^{(k)}_{eff}(\overline
      f+f)},
\label{Cell-Pint}
\end{eqnarray}
where $S^{(k)}_{eff}[f]$ is the effective action of the kth cell and the path
integration of $f$ is over those ones that equal to the mean field value
$\overline f$ outside of the kth cell but with arbitrary amplitudes inside the
finite volume. For a given theory, instead of arbitrary division of the
spacetime, it is expected that there is an optimal one with minimum volume
$\varpi$ for each cell and yet has an error below a predetermined one. We
shall assume that such an optimal division of the spacetime into cells has
already been found in the following discussion.

Let us evaluate the contributions of the uncorrelated localized quantum
fluctuations of the $\sigma$ and $\vec{\pi}$ fields to the vacuum fermion
number density using Eqs. \ref{rho-general}, \ref{Cluster-Decomp} and
\ref{Cell-Pint} in the Euclidean spacetime.

In the phase where the chiral $SU(2)_L\times SU(2)_R$ symmetry is
spontaneously broken down, the ``chiral angle'' variable represented by
$\vec{\pi}$ in the phase where $ \mathopen{\langle 0\,|} \vec{\pi}
\mathclose{|\,0 \rangle}=0$ becomes massless following the Goldstone theorem.
The correlation length of the $\vec{\pi}$ field becomes divergent in the
chiral symmetric limit. Therefore the field configurations of the Goldstone
boson degrees of freedom contain the dominating on-shell components that can
be included by doing a loop expansion as usual. Such a loop expansion
contains no infrared divergences. Because the fermion number density $\rho
(\sigma,\vec{\pi},\mu;x)$ under study is chiral symmetric, which means that
it does not depend on a spacetime independent global ``chiral angle''; it
depends only on the derivatives of the ``chiral angle'' variables. The
absence of a dependence  of the fermion number density on a global ``chiral
angle'' guarantees the absence of the infrared divergences in the quantum
corrections from the Goldstone bosons. The configurations of the ``chiral
angle'', being on the edge of their shell and extended configurations in
nature, are uniform and infinitesimal in amplitudes since it contains no
infrared divergences and has a number of distinct modes proportional to the
volume $\Omega $ of the system. They can not modify the qualitative features
of the quasiparticles. So, the $\vec{\pi}$ variable in the vacuum fermion
number density can be eliminated. It is treated as zero in the following
discussions.

The quantum fluctuation in the ``mass'' term, namely the chiral radius or
order parameter represented by $\sigma$ (when $ \mathopen{\langle
0\,|} \vec{\pi} \mathclose{|\,0\rangle}=0$) has different 
characteristics due to the fact that it contains no on-shell Euclidean 
configurations. These
off-shell configurations have only short range correlations in spacetime.

   If the spacetime is divided into cells with their dimension optimally
determined, then the path integration within each cell can be done
independently. This gives us a cluster decomposed partition functional of the
form given by Eq. \ref{Cluster-Decomp} with the partition functional for each
cell computed independent of each other.

The cluster decomposition property of the partition functional of the system
reduces the full fermion number density of the vacuum given by Euclidean form
of Eq. \ref{rho-general} to
\begin{eqnarray}
       \rho_{vac}  &=& {1\over
       z_k[0]}\int D_k[\sigma'] \rho[\overline\sigma+\sigma',0,\mu;x]
       e^{-S^{(k)}_{eff}[\overline\sigma+\sigma',\mu]}, \label{rho-decomp}
\end{eqnarray}
where the cell labeled by k is the one that contains the spacetime point $x$,
$\int D_k[\sigma']$ denotes integration over field configurations that
approach the mean field value outside the cell and $S^{(k)}_{eff}$ is the
Euclidean effective action of the cell.

Eq. \ref{rho-decomp} is still too complicated to evaluate analytically. We
make a further simplification by assuming that the functional integration of
$\sigma'$ within a spacetime cell can be replaced by an ordinary integration
over the spacetime averaged value of $\sigma'$ within that cell. It can be
achieved by keeping the average of $\sigma'$ fixed while integrate over the
rest degrees of freedom. The non-trivial part of it is in the assumption that
after eliminating the rest of the degrees of freedom, the resulting
effective potential $V_{eff}$ remains, at least in form, the same as the
original one. This procedure is in the same spirit as the renormalization
group analysis. In this way, Eq. \ref{rho-decomp} reduces to
\begin{eqnarray}
       \rho_{vac} &=& {1\over
       z[0]}\int^\infty_{-\infty}
       d\delta\sigma \rho[\overline\sigma+\delta\sigma,0,\mu;x]
       e^{-\varpi V_{eff}(\overline\sigma+\delta\sigma,\mu)},
\label{rho-decomp1}
\end{eqnarray}
where $\delta\sigma=<\sigma'>$ is the spacetime average of $\sigma'$ within
the cell and the same reduction is also made to $z[0]$. The finiteness of
$\varpi $ result in different qualitative behavior for $ \rho_{vac}$ as a
function of $\mu$. To explicitly see the difference, let us expand $V_{eff}$
around $\overline\sigma$, keeping only the leading quadratic term
\begin{eqnarray}
   V_{eff}(\overline\sigma+\delta\sigma) &=& V_{eff}(\overline\sigma) +
                              A(\overline\sigma) \delta\sigma^2 + \ldots
\end{eqnarray}
and using Eq. \ref{rhoMF} for the trace term. If $\varpi A$ is sufficiently
large, the result can be written as
\begin{eqnarray}
\rho_{vac} &=& {2\over \pi^2} \sqrt{\varpi
 A\over\pi}\int_{-\mu-\overline\sigma}^{\mu-\overline\sigma} d\delta\sigma
        e^{-\varpi A \delta\sigma^2} \left
        [\mu^2-(\overline\sigma+\delta\sigma)^2\right ]^{3/2}.
\end{eqnarray}
It is clearly non-zero for any finite $\mu$ below $\overline\sigma$
since $A$ is finite. The $\mu$ dependence of $\rho_{vac}$ differs from
the form given by Eq. \ref{rhoMF} as long as $G_0$ is
finite. Eq. \ref{rhoMF} is recovered only in the limit of $G_0\to 0$,
which represents the free field case. The dependence of
$\rho_{vac}^{1/3}$ on $\mu$ for a set of different values of
$\alpha=\varpi A$ is plotted in Fig. \ref{Fig:Fdensity}. Instead of a sharp
rise in $\rho^{1/3}_{vac}$ at $\mu = \overline\sigma$, $\rho$ is
non-vanishing all the way to $\mu=0$. This component of the fermion
number density can certainly not be attributed to the contributions of
the quasiparticles. It is found therefore that there is a dark
component for the fermion number density that can not be accounted for
by the quasiparticle contributions.

The NJL model has only one non-trivial vacuum phase since there is one
independent absolute minimum in its effective potential. If the effective
potential of the system contains a second local minimum with a higher energy
density than the absolute minimum, then there are contributions from the local
minimum to the fermion number density. The existence of such a contribution
is another direct consequences of the cluster decomposability of the
partition functional of the system. Models with a second minimum are studied
in the literature. Those that have only one order parameter are represented
by the Friedberg--Lee model \cite{FL-model} in which the two minima of the
effective potential correspond to confinement and deconfinement phase of the
model. Those that has two order parameters are introduced in section 
\ref{sec:Models}.

In the presence of a higher virtual phase that is separated from the actual
phase of the system by a potential barrier, the fermion number density of the
system is saturated by both the quasiparticles of the first phase and those
ones of the second virtual phase together with their corresponding dark
component discussed above. It has a form
\begin{eqnarray}
\rho_{vac} &=& {\rho_{vac_1} + e^{-\varpi
  \Delta} \rho_{vac_2}\over 1 + e^{-\varpi \Delta}},
\label{2phase-rho}
\end{eqnarray}
where $\Delta=V_{eff}^{(2)}-V_{eff}^{(1)}>0$ is the difference in energy
density between the virtual phase and the actual phase, $\rho_{vac_1}$ and
$\rho_{vac_2}$ are the vacuum fermion number density of the actual vacuum
phase and that of the virtual vacuum phase respectively and $\varpi$ is the
optimal volume of the spacetime cell between which the order parameters of
the system are discorrelated. The contributions of the virtual phase to the
fermion number density or to any local physical observables are
non-perturbative effects. Attempts had been made in Refs.
\cite{PCACsuc,GDHp,pPaps1} to search for possible other virtual phases
like the $\beta$ or $\omega$ phase of the strong interaction vacuum.

\subsection{The reemergence of the quasiparticle picture}

   The fermion number density discussed above are defined on a
spacetime point. Such a precision is non-achievable in realistic
observations. The physical observables can be represented by a
``coarse-grained averaging'' of the form $ \overline\rho_{vac}={\Delta
N_{vac} /\Delta\Omega}$ with $\Delta\Omega$ the smallest volume in
spacetime that the observation apparatus can resolve and $\Delta
N_{vac}$ the average number of fermions due to the {\em coherent}
response of the system to an external classical field $K$ within that
volume. $\Delta N_{vac}$ is not in general identical to
$\int_{\Delta\Omega} d^4 x \rho_{vac}$ when $\Delta\Omega>>\varpi$. It
is obtained from the partition functional $W[J,\mu]$ by adding the
external field $K$ to $\mu$.  $K$ has a constant non-zero value only
within spacetime volume $\Delta
\Omega$, then $\Delta N_{vac} = \partial W[0,\mu+K]/\partial iK |_{K=0}$. In
case when $\Delta \Omega << \varpi$, namely the precision of the
observation is much higher than the correlation length of the order
parameter, the observed fermion number density $\overline\rho_{vac}$
behaves in the same way as $\rho_{vac}$. On the other hand, if
$\Delta\Omega >> \varpi$, then the smallest spacetime cell that
contributes to $\Delta N_{vac}$ is a region of volume $\Delta\Omega$
rather than $\varpi$. In this case, $\overline\rho_{vac}$ is obtained
from $\rho_{vac}$ by substituting $\Delta\Omega$ for $\varpi$. The
effects of the dark component in the observed fermion number density
are reduced as a result. When the resolution is sufficiently low,
which means $\Delta \Omega A >> 1$, the quasiparticle picture with
only one vacuum , namely the actual one, reemerges in the response of
the system to $K$.  Therefore the dark component of the fermion number
density is of transient nature. In most of the situations encountered
in non-relativistic condensed matter system, the condition $\Omega A
>> 1$ is satisfied so that a global chemical potential approach is
sufficient.

\subsection{The observation of the dark component at low energies}

On the other hand, the $K$ field radiated by the vacuum fermion number density
is of the form $K(x)=\int d^4 x'G(x,x')\rho_{vac}(x')$ with $G(x,x')$ the 
Green function for $K(x)$. In a local QFT, 
it is $\rho_{vac}$ that is the source for $K(x)$ 
rather than $\overline\rho_{vac}$ no matter how slow the resulting $K(x)$ 
varies in spacetime. Thus the effects of the dark component are indirectly 
observable even in low energy processes.

\section{QCD Lagrangian Density with Quark Field in the Real 
                    8 Component Representation}
\label{app:QCD}

The QCD Lagrangian density for strong interaction can be written as
\begin{eqnarray}
    {\cal L}_{QCD} & = & -{1\over 2} Tr G^{\mu\nu} G_{\mu\nu} +
           {1\over 2}\overline\Psi\left (
            i\rlap\slash\partial - i g\hspace{0.06cm}\rlap\slash
           \hspace{-0.06cm} A - m_0 \right ) \Psi,
\label{QCD1}
\end{eqnarray}
where
\begin{eqnarray}
            G^{\mu\nu} = \partial^\mu A^\nu - \partial^\nu A^\mu
             - ig[A^\mu,A^\nu],&\hspace{0.8cm}&
               A^\mu = \sum_{a=1}^8 {\cal B }^\mu_a {\Lambda^a\over 2}
         \label{QCD2}
\end{eqnarray}
and
\begin{eqnarray}
         \Lambda^a &=& {1\over 2} \left
             [(1+O_3)\lambda^a - (1-O_3)\lambda^{aT} \right ]\label{QCD3}
\end{eqnarray}
with $m_0$ current quark mass, $g$ the QCD coupling constant, ${\cal
  B}^a$ the bosonic gauge fields,
``T'' representing transpose and
$\lambda^a$ $(a=1,2,\ldots,8)$ the Gell-Mann matrices.

  It is easy to verify that $\Lambda^a$ $(a=1,2,\ldots,8)$
satisfy the same commutation relation with each other as the
Gell--Mann matrices $\lambda^a$ do, namely,
\begin{eqnarray}
     \left [\Lambda^a,\Lambda^b \right ] &=& if^{abc}\Lambda^c,
\label{SU3-comm}
\end{eqnarray}
where $f^{abc}$ (with $a,b,c=1,2,\ldots,8$) are the set of group structure
constants of $SU(3)$. Therefore $\Lambda^a$ belong to the same adjoint
representation of $SU(3)$ as the one that $\lambda^a$ belong.

\section{A Formal Connection Between $S_{\lowercase{eff}}[\lowercase{f}]$ and 
        \lowercase{$\Gamma[f]$}}
\label{app:CJT}

A formal approach to relate $S_{eff}[f]$ and $\Gamma[f]$ can be
found using the method developed in Ref. \cite{Jackiw} by identifying
$S_{eff}[f]$ as $I[\phi]$. Assuming that $f$ are all real, which does not loss
any generality since a complex field can be regarded as two real
fields, the relation between $\Gamma[f]$ and $S_{eff}[f]$ can be 
established through a generalized vertex functional
$\widetilde\Gamma[f,G]$, which is  defined by
\begin{eqnarray}
        \widetilde\Gamma[f,G] &=& iS_{eff}[f] - {1\over 2} Sp Ln D G^{-1}
                        - {1\over 2} Sp \left (
                       {\cal D}^{-1} G - 1 \right ) + \Gamma_2[f,G],
\label{G-S-rel-M}
\end{eqnarray}
where $\Gamma_2[f,G]$ is the contributions of all two particle
irreducible graphs with lines representing $G$ in the shifted
background fields $f$ and vertices obtainable from $S_{eff}[f]$ by
expanding it around a set of shifted $\{ f_i \}$ and keeping terms cubic in
$f$ or higher. Here $D$ is the bare propagator for the boson
fields and ${\cal D}^{-1}$ is symbolically given by
\begin{eqnarray}
    {\cal D}^{-1} &=& {\delta^2 S_{eff}[f] \over \delta f \delta f}.
\end{eqnarray}
The proper vertex function $\Gamma[f]$ equals to $\widetilde\Gamma[f,\overline G]$ with
$\overline G$ satisfying the equation
\begin{eqnarray}
       \left . {\delta \widetilde\Gamma[f,G] \over \delta G} \right |_{G=\overline G}
        &=& 0.
\end{eqnarray}

In the Euclidean spacetime formulation, Eq. \ref{G-S-rel-M} becomes
\begin{eqnarray}
        \widetilde\Gamma[f,G] &=& S_{eff}[f] - {1\over 2} Sp Ln D G^{-1}
                        - {1\over 2} Sp \left (
                       {\cal D}^{-1} G - 1 \right ) + \Gamma_2[f,G],
\label{G-S-rel-E}
\end{eqnarray}
where all the quantities above are evaluated in the Euclidean spacetime.

\section{The Nambu Jona--Lasinio Model}
\label{app:NJL}

For the simplicity of the discussion, I consider a two flavor NJL
model \cite{NJL}. In the conventional 4-dimensional representation for the
fermion fields, it takes the following form
\begin{eqnarray}
     {\cal L} &=& \overline\psi i\rlap\slash\partial \psi +  G_0\left [
      \left (\overline\psi\psi\right)^2 + \left (\overline\psi
      i\gamma^5\vec{\tau}\psi\right )^2 \right ]
\label{Model-L}
\end{eqnarray}
with $G_0$ the coupling constant and $\vec{\tau}$ the three
Pauli matrices in the isospin space. It has a $SU(2)_L\times SU(2)_R$ chiral
symmetry. Due to the non-linear 4-fermion interaction term, it is not
directly solvable. One of the best way to tackle the non-linear
4-fermion interaction model is to introduce auxiliary fields
\cite{Ying11}. For this model two sets of auxiliary fields, namely $\sigma$ and
$\vec{\pi}$ are necessary. The model Lagrangian density after the introduction
of
the auxiliary fields $\sigma$ and $\vec{\pi}$ is of the following form
\begin{eqnarray}
   {\cal L} &=& {1\over 2}\overline\Psi\left (i\rlap\slash\partial - \sigma -
   i \vec{\pi}\cdot\vec{\tau} \gamma^5 O_3  \right ) \Psi -
     {1\over 4 G_0} \left (\sigma^2 + \pi^2
 \right ),
\label{Model-L2}
\end{eqnarray}
which is written in terms of the 8-dimensional representation $\Psi$ for
the Dirac spinors for fermions.

The effective potential for this model can be computed using the
Euclidean contour shown in Fig. \ref{Fig:ConCl2}. It has the
following form
\begin{eqnarray}
    V^{eff} (\sigma^2) &=& {\Lambda^4\over 4\pi}\left [
{1\over 4}\left ({1\over \alpha_0} - {6\over\pi}\right
){\sigma^2\over
\Lambda^2} + {3\over 2\pi}
\ln\left ( 1+{\Lambda^2\over \sigma^2}\right )
{\sigma^4\over\Lambda^4}
- {3\over 2\pi} \ln\left (1+{\sigma^2\over
\Lambda^2} \right )\right ]
\label{Veff-NJL}
\end{eqnarray}
with $\alpha_0 = G_0\Lambda^2/4\pi$.

It has two phases. The system is in the first one, namely the $O$
phase,  when $\alpha_0\le
\pi/12$. It is in the second one called the {\em $\alpha$ phase} 
when the coupling constant $\alpha_0>\pi/12$, where the original chiral
$SU(2)_L\times SU(2)_R$ symmetry is spontaneously broken down to a
$SU(2)_V$ flavor (isospin) symmetry and quark--antiquark pair
condenses. Within the approximation adopted, the phase
transition across the $\alpha_0=\pi/12$ point is second order.

\end{appendix}

\end{document}